\documentstyle[eqsecnum,aps]{revtex}

\begin{document}
\draft

\newcommand{\BQ}{\begin{equation}}
\newcommand{\EQ}{\end{equation}}
\newcommand{\BQA}{\begin{eqnarray}}
\newcommand{\EQA}{\end{eqnarray}}
\newcommand{\half}{\frac{1}{2}}
\newcommand{\NN}{\nonumber \\}
\newcommand{\E}{{\rm e}}
\newcommand{\Gmu}{\gamma^{\mu}}
\newcommand{\Gnu}{\gamma^{\nu}}
\newcommand{\gmu}{\gamma_{\mu}}
\newcommand{\gnu}{\gamma_{\nu}}
\newcommand{\gfive}{\gamma_5}
\newcommand{\del}{\partial}
\newcommand{\k}{\mbox{\boldmath $k$}}
\newcommand{\itl}{\mbox{\boldmath $l$}}
\newcommand{\itP}{\mbox{\boldmath $P$}}
\newcommand{\q}{\mbox{\boldmath $q$}}
\newcommand{\p}{\mbox{\boldmath $p$}}
\newcommand{\x}{\mbox{\boldmath $x$}}
\newcommand{\y}{\mbox{\boldmath $y$}}
\newcommand{\Z}{{\bf Z}}
\newcommand{\R}{{\rm R}}
\newcommand{\M}{{\cal M}}
\newcommand{\T}{{\cal T}}
\newcommand{\U}{{\cal U}}
\newcommand{\SS}{{\cal S}}
\newcommand{\PP}{{\cal P}}
\newcommand{\tr}{{\rm tr}}
\newcommand{\Path}{{\rm P}\,}
\newcommand{\dagg}{\mbox{\scriptsize{\dag}}}
\newcommand{\dagpsi}{\psi^{\mbox{\scriptsize{\dag}}}}
\newcommand{\ket}[1]{\left.\left\vert #1 \right. \right\rangle}
\newcommand{\bra}[1]{\left\langle\left. #1 \right\vert\right.}
\newcommand{\ketrm}[1]{\vert {\rm #1} \rangle}  
\newcommand{\brarm}[1]{\langle {\rm #1} \vert}  
\newlength{\bredde}
\def\slash#1{\settowidth{\bredde}{$#1$}\ifmmode\,\raisebox{.15ex}{/}
\hspace*{-\bredde} #1\else$\,\raisebox{.15ex}{/}\hspace*{-\bredde} #1$\fi}
\renewcommand{\thefootnote}{\fnsymbol{footnote}}


\title{Dynamical Chiral Symmetry Breaking on the Light Front.\  II.\\   
   The Nambu--Jona-Lasinio Model} 
\author{K. Itakura\footnote{Address after June 1st: RIKEN-BNL Research Center, BNL, Upton, NY 11973, USA.}}
\address{Research Center for Nuclear Physics, Osaka University, 
Osaka 567-0047, Japan\\
{\tt itakura@rcnp.osaka-u.ac.jp}}
\author{S. Maedan}
\address{Department of Physics, Tokyo National College of Technology, \\
Tokyo 193-8610, Japan\\
{\tt maedan@tokyo-ct.ac.jp}}
\maketitle

\begin{abstract}
An investigation of dynamical chiral symmetry breaking 
 on the light front is made in the Nambu--Jona-Lasinio model 
 with one flavor and $N$ colors. 
Analysis of the model suffers from extraordinary complexity due to the 
  existence of a ``fermionic constraint," i.e., a constraint equation
  for the bad spinor component.
However, to solve this constraint is of special importance.
In classical theory, we can exactly solve it and 
  then explicitly check the property of ``light-front chiral transformation.''
In quantum theory, we introduce a bilocal formulation to solve the 
  fermionic constraint by the $1/N$ expansion.
Systematic $1/N$ expansion of the fermion bilocal operator is realized by
  the boson expansion method. 
The leading (bilocal) fermionic constraint becomes a gap 
  equation for a chiral condensate and thus if we choose a nontrivial 
  solution of the gap equation, we are in the broken phase.
As a result of the nonzero chiral condensate, we find unusual chiral 
  transformation of fields and nonvanishing of the light-front 
   chiral charge.
A leading order eigenvalue equation for a single bosonic state is 
  equivalent to a leading order fermion-antifermion bound-state equation. 
We analytically solve it for scalar and pseudoscalar mesons and obtain 
   their light-cone wavefunctions and masses.
All of the results are entirely consistent with those of our previous 
   analysis on the chiral Yukawa model.
\end{abstract}


%
%
\section{Introduction}

Our expectation that light-front (LF) formalism enables us to 
  relate QCD directly to the constituent quark model at field-theoretic 
  level~\cite{Wilson}, seriously requires a full understanding 
  of dynamical chiral symmetry breaking (D$\chi$SB) on the LF.
Central to this issue is, first of all, a well-known problem of 
  how to reconcile a LF ``trivial'' vacuum with a chirally broken 
  vacuum having a nonzero fermion condensate.
The secondary problem is to determine the property of ``LF
  chiral transformation'' which is defined differently from the usual one.
The most surprising fact of the LF chiral transformation is that it is an 
  exact symmetry even for a massive free fermion~\cite{Mustaki}.

In the present paper, we discuss this issue within the Nambu--Jona-Lasinio 
  (NJL) model~\cite{NJL} which is a typical example of D$\chi$SB.
Previously we consider the same problem from different point of view~\cite{I}.
Our interest was in describing D$\chi$SB of the NJL model, 
  but we actually took a roundabout way in order to apply an idea 
  which works well for spontaneous symmetry breaking of 
  a scalar model, to a fermionic theory.
We know that the longitudinal zero modes of scalar fields are responsible 
  for describing spontaneous symmetry breaking on the LF.
Indeed, it is achieved by solving the ``zero-mode constraints''
  (i.e., constraint equations for the longitudinal zero modes)
  nonperturbatively~\cite{SSB}.
The zero-mode constraint appears in the discretized light-cone quantization 
  (DLCQ) approach~\cite{DLCQ}, where we set periodic boundary conditions 
  for scalars in the longitudinal direction with finite extension. 
Of course the NJL model has no scalar fields as fundamental 
  degrees of freedom, but we overcame the situation by considering  
  the chiral Yukawa model.
This model shows D$\chi$SB in large $N$ limit ($N$ is 
  the number of fermions) and goes to the NJL model in infinitely 
  heavy mass limit of scalar and pseudoscalar bosons.
We showed that the zero-mode constraint of the scalar field correctly 
  produces a gap equation for a chiral condensate and calculated 
  masses of the scalar and pseudoscalar bosons from poles of their propagators.
Therefore, in Ref.~\cite{I}, we succeeded in describing 
  {\it indirectly} the chiral symmetry breaking of the NJL model on the LF. 

Since the very essence of the previous analysis was the existence 
  of scalar fields, one may ask a question:  How can one formulate 
  D$\chi$SB {\it without} scalars?
In order to answer the question, we treat the NJL model 
  without introducing auxiliary field.
An important key was already shown in Ref.~\cite{Itakura}.
It was found that the ``fermionic constraint'' plays the same role 
  as that of the zero-mode constraint.
Splitting the fermion field as 
  $\Psi=\psi_++\psi_-,\ \psi_\pm=\Lambda_\pm\Psi$ by using projectors 
  $\Lambda_\pm=\gamma^\mp\gamma^\pm/2$, we easily find that the 
  ``bad'' component $\psi_-$ is a dependent variable and subject 
  to a constraint equation called the ``fermionic constraint.''
In the LF NJL model, the fermionic constraint is very complicated and 
 it is difficult to solve it as an operator equation.
However, we will see that to solve this equation is a crucial step
  for describing the broken phase and will find a close parallel between 
  the fermionic constraint for D$\chi$SB and the zero-mode constraint 
  for spontaneous symmetry breaking in scalar models.
Although such special importance of the fermionic constraint might 
  be restricted only to the LF NJL model and therefore most of the 
  analysis might be model dependent, but what we are eventually 
  interested in is the physics consequences of the chiral symmetry breaking. 
And of course we cannot reach the chiral symmetry breaking in QCD 
 unless we understand the simpler and typical example of this phenomenon.
Therefore the importance of our analysis is evident.

Let us comment on other attempts for the NJL model on the LF.
First of all, Heinzl et al.~\cite{Heinzl} treated the model within 
  the mean-field approximation and insisted delicacy of the 
  infrared cutoff to obtain a chiral condensate.
The meaning and necessity of such cutoff scheme was clarified 
  in Ref.~\cite{ItakuraMaedan}. 
As mentioned above, an observation that a gap equation for a chiral 
  condensate emerges from the fermionic constraint was first pointed 
  out by one of the authors~\cite{Itakura}.
The light-cone (LC) wavefunction of a pionic state was calculated 
  through the LC projection of the Bethe-Salpeter amplitude 
  which was derived in the equal-time quantization~\cite{Heinzl_review}.
Bentz et al. introduced the auxiliary fields to fermion bilinears and 
  solved the constraint equations for them by $1/N$ expansion~\cite{Bentz}.
They obtained ``effective'' Lagrangian for the broken phase and discussed 
  the structure function of the pionic state.
With all these studies, however, there still remains many unknowns 
  concerning basic problems.
Especially, we still do not understand well the LF chiral transformation 
  itself.
To what extent is it different from the usual chiral symmetry? 
How is the chiral symmetry breaking realized on the LF?
These fundamental problems will be resolved in the present paper.

The paper is organized as follows. 
The rest of this section is devoted to introduction of the NJL model 
  and our notation.
In  Sec.~II, we discuss the complexity of the fermionic constraint
  in great detail. 
We explicitly solve the fermionic constraint in classical treatment and 
  investigate properties of LF chiral transformation. 
In Section III, we solve the fermionic constraint in quantum theory 
  by the $1/N$ expansion.
Here we introduce the boson expansion method in order to solve the 
  bilocal fermionic constraint with systematic $1/N$ expansion.
We see emergence of the gap equation for the chiral condensate 
  from the fermionic constraint.
We obtain the Hamiltonian with respect to the (bilocal) bosons which 
  is introduced by the boson expansion method.
In Section IV, some physics consequences of the chiral symmetry breaking are 
  discussed. 
First of all, we see how the chiral symmetry breaking is realized 
  in the LF formalism.
We discuss unusual chiral transformation of fields and nonconservation 
  of the light-front chiral charge.
Secondly, we construct the bound-state equation for mesonic states and 
  solve it for scalar and pseudoscalar mesons.
Thirdly, we derive the partially conserving axial current (PCAC) relation.
Summary and conclusion are given in the last section.
Miscellaneous topics with detailed calculation are presented in Appendices.

Before ending this section, let us fix our model and notation.
Since the primary purpose of our paper is to study basic properties 
  of the LF chiral symmetry, we consider only one flavor case 
  for simplicity.
Thus the model we discuss is 
\BQ
{\cal L}=\bar\Psi(i\slash\del -m_0)\Psi +
\frac{g^2}{2}\left[(\bar\Psi\Psi)^2+(\bar\Psi i \gamma_5 \Psi)^2\right]~.
\EQ
Here  $\Psi_\alpha^a(x)  \  (a=1,\ldots,N)$ is a four component spinor 
   with ``color'' internal symmetry $U(N)$, which has been introduced 
   so that we can use the $1/N$ expansion as a nonperturbative technique.
We always work with a nonzero bare mass $m_0\neq 0$.
The primary reason is that the Hamiltonian with a massless fermion 
  is plagued with a troublesome situation in 1+1 dimensions:
As we will see, if we set $m_0=0$ from the beginning, the canonical LF 
  Hamiltonian $P^-$ of the Gross-Neveu model vanishes at all.
Even in 3+1 dimensions, we will see that absence of the bare mass term 
  causes an inconsistency of the results. 
The secondary reason is to avoid massless particles which can move in 
  parallel with $x^+=$constant surface.
The difficulty of describing massless particles is intimately connected with 
  the fact that on the LF, the (massless) Nambu-Goldstone 
  boson becomes physically meaningful only when we first 
  include explicit breaking term and then take the vanishing 
  limit of it~\cite{Tsujimaru-Yamawaki}.
The same situation was observed in the chiral Yukawa model~\cite{I}.

In practical calculation, it is convenient to introduce the two-component
  representation for the gamma matrices so that the projectors $\Lambda_\pm$ 
  are expressed as
\BQ
\Lambda_+=\pmatrix{
{\bf 1} & 0 \cr
0 & 0 \cr
},\quad
\Lambda_-=\pmatrix{
0 & 0 \cr
0 & {\bf 1} \cr
}.\label{two-compo}
\EQ
Then, the projected fermions have only upper or lower components:
\BQ
\psi_+=\! \Lambda_+\Psi\equiv 2^{-\frac{1}{4}}\left (\matrix{
\psi \cr
0\cr
}\right ),\quad 
\psi_-=\! \Lambda_-\Psi\equiv 2^{-\frac{1}{4}}
\left (\matrix{
0\cr
\chi \cr
}\right ),\label{spinors}
\EQ
where we defined two-component spinors $\psi$ and $\chi$.
Among various representations which satisfy Eq.~$(\ref{two-compo})$, 
  we choose a representation having a similar structure 
  to the chiral representation in 1+1 dimensions, i.e., $\gamma^0=\sigma^1,\ 
  \gamma^1=-i\sigma^2,\ \gfive=\gamma^0\gamma^1=\sigma^3$.
Explicitly, they are 
\BQ
\gamma^0=\pmatrix{
0 & {\bf 1}\cr
{\bf 1} & 0\cr
},\quad 
\gamma^3=
\pmatrix{
0 & -{\bf 1} \cr
{\bf 1} & 0 \cr
},\quad
\gamma^i=
\pmatrix{
-i\sigma^i & 0 \cr
0 & i\sigma^i \cr
},\label{two-compo-gamma}
\EQ
for $i=1,2$ and 
\BQ
\gamma_5=\pmatrix{
\sigma^3 & 0 \cr
0 & -\sigma^3 \cr
}.\label{gammafive}
\EQ
Results of the chiral Gross-Neveu model in 1+1 dimensions can be 
  easily obtained if we make a replacement for the Pauli matrices 
  $\sigma^3 \rightarrow 1$ and $\sigma^i\rightarrow 0$, 
  and regard $\psi$ and $\chi$ as one component spinors. 
The explicit form of the Lagrangian in the two-component representation 
  is given in Appendix A.

In the previous work~\cite{I}, we made the longitudinal direction finite 
  in order to carefully treat the longitudinal zero modes of the scalar fields.
However, in the present analysis, we work in an infinite longitudinal space. 
There is no need of introducing finite $x^-$. 
When we take the inverse of $\del_-$, we need to specify the boundary 
  conditions.
We here follow the conventional antiperiodic boundary 
  condition\footnote{For a scalar field, antiperiodic boundary condition in 
  infinite longitudinal space leads to inconsistency~\cite{Tsujimaru-Yamawaki}.
  However, fermionic fields are free from such troubles.} 
  which is standard for free fermions:
\BQA
\psi_a(x^-=-\infty, x^i_\perp)&=&-\psi_a(x^-=\infty, x^i_\perp)~,\\
\chi_a(x^-=-\infty, x^i_\perp)&=&-\chi_a(x^-=\infty, x^i_\perp)~.
\EQA
So our results must always have a smooth free field limit.

%
%
\setcounter{equation}{0}
\section{Complexity of the fermionic constraint and the chiral symmetry}

It is highly complicated structure of the fermionic constraint 
  that makes the analysis of the LF NJL model difficult.
However, we cannot know anything about LF chiral symmetry unless 
  we confront with this complexity.
Therefore in this section, we investigate the fermionic constraint 
  in great detail.
First of all, we classically solve the fermionic constraint.
Using the explicit solution, we then discuss properties of the LF 
  chiral transformation.
Especially we show that LF chiral transformation is no longer an exact 
  symmetry when $m_0 \neq 0$.
Finally we consider the implication of the fermionic constraint 
  in quantum theory.

\subsection{Fermionic constraint and its classical solution}

The fermionic constraint is immediately obtained as the
  Euler-Lagrange equation for $\chi$:
\BQ
i\del_-\chi_a=\frac{1}{\sqrt{2}}\left(-\sigma^i\del_i+m_0\right)\psi_a
-\frac{g^2}{2}\left\{
   \psi_a\left(\psi^\dagger_b\chi_b+\chi^\dagger_b\psi_b\right)+
   \sigma^3\psi_a\left(\psi^\dagger_b\sigma^3\chi_b
                    -\chi^\dagger_b\sigma^3\psi_b\right)
\right\}~,\label{FC}
\EQ
where summation over color and spinor indices are implied.
If we want to solve this equation as an operator equation in quantum theory, 
 we need a commutation relation between $\chi$ with $\psi$ which must 
 be given by the Dirac bracket.
Since the anticommutator $\{\chi, \psi\}$ is very complicated, 
 it seems almost hopeless to find an exact quantum solution of it.
However, in a classical theory where we treat all the variables just 
 as Grassmann numbers, the equation becomes tractable
 and it is not difficult to solve it. 
Indeed, the exact solution with antiperiodic boundary condition 
 is given by (see Appendix B for more details)
\BQ
\pmatrix{
\chi_{1a}(x)\cr
-\chi_{2a}^\dagger(x)
}=\frac{1}{\sqrt2}\int_{-\infty}^\infty dy^- G_{ab}(x^-,y^-,x_\perp)
\pmatrix{
m_0\psi_{1b}(y^-)-\del_z\psi_{2b}(y^-)\cr
-\del_z\psi_{1b}^\dagger(y^-)+m_0\psi_{2b}^\dagger(y^-)
},\label{classical_solution}
\EQ
where $\del_z=\del_1-i\del_2$ and 
the ``Green function'' $G_{ab}(x^-,y^-,x_\perp)$ is 
\BQA
&&G_{ab}(x^-,y^-,x_\perp)
=G^{(0)}(x)\left[\ \frac{1}{2i}\epsilon(x^--y^-)+C\ \right]G^{(0)-1}(y)~,
\label{Green_NJL}\\
&&\quad G^{(0)}(x)={\rm P}\  e^{ig^2\int_{-\infty}^{x^-}{\cal A}(y^-)dy^-}~,
\label{Path}\\
&&\quad {\cal A}_{ijab}=\pmatrix{
\psi_{1a}\psi_{1b}^\dagger & \psi_{1a}\psi_{2b} \cr
\psi_{2a}^\dagger\psi_{1b}^\dagger & \psi_{2a}^\dagger \psi_{2b} \cr
}~.\label{matrix_A}
\EQA
The integral constant $C$ is determined so that the solution satisfies 
  the antiperiodic boundary condition.
In Eq.~(\ref{Path}), P stands for the path-ordered product.
Note that  we easily derive the solution for the chiral Gross-Neveu
model by extracting the 1-1 component of ${\cal A}$ and neglecting
$\del_z$.
The result is equivalent to the solution of the Thirring model 
 obtained by Domokos~\cite{Domokos}.
And also, if we take the free fermion limit $g^2\rightarrow 0$, 
  we of course recover the free solution due to $G^{(0)}(x)\rightarrow 1$ 
  and $G(x^-,y^-)\rightarrow \epsilon(x^--y^-)/2i$.

\subsection{Chiral symmetry on the light front}
Since the bad component $\chi$ is a constrained variable in the LF formalism,
we impose the chiral transformation only on the 
good component $\psi_+\longrightarrow e^{i\theta \gfive }\psi_+$
or in the two-component representation [see Eq.~(\ref{gammafive})]
\BQ
\psi\longrightarrow e^{i\theta \sigma^3}\psi~. \label{Chiral_Transf}
\EQ
Now we have completely solved the fermionic constraint for $\chi$, 
  we can explicitly demonstrate its transformation property under the LF 
  chiral transformation.
However, before discussing the NJL model, it will be instructive to remind 
  you of the LF chiral symmetry in the free massive fermion.

As we mentioned before, the {\it massive}  free fermion is chiral invariant 
  under the transformation (\ref{Chiral_Transf}). 
Let us see this fact directly in the Lagrangian even though it is a little 
  lengthy.
It is convenient to separate the solution of the fermionic constraint 
  $\chi=(\sqrt{2}i\del_-)^{-1}\left(-\sigma^i\del_i+m_0\right)\psi$ 
  into mass-independent and dependent parts $\chi=\chi^{(0)} + \chi^{(m)}$ 
  as
$$
\chi^{(0)}=-\frac{1}{\sqrt2}\sigma^i \del_i \frac{1}{i\del_-}\psi~,
\qquad \chi^{(m)}=\frac{m_0}{\sqrt2}\frac{1}{i\del_-}\psi~.
$$
Note that there is a relation between $\chi^{(0)}$ and $\chi^{(m)}$:
\BQ
\sigma^i \del_i \chi^{(m)}+m_0\chi^{(0)}=0~.\label{relation}
\EQ
As a result of the LF chiral transformation (\ref{Chiral_Transf}), 
we find
\BQA
&&\chi^{(0)}\longrightarrow e^{-i\theta\sigma^3}\chi^{(0)}~,\label{transf0}\\
&&\chi^{(m)}\longrightarrow e^{i\theta\sigma^3}\chi^{(m)} ~.\label{transfm}
\EQA
The free fermion Lagrangian is compactly expressed as
${\cal L}_{\rm free}= \psi^\dagger \omega_{\rm EOM}
                     +\chi^\dagger\omega_{\rm FC},$ 
where 
  $\omega_{\rm EOM}=i\del_+\psi 
   -\frac{1}{\sqrt2}(\sigma^i\del_i+m_0)\chi=0$ 
  is the equation of motion for $\psi$ and 
  $\omega_{\rm FC}=i\del_-\chi -\frac{1}{\sqrt2}(-\sigma^i\del_i+m_0)\psi=0$ 
  is the fermionic constraint.
The second term is zero and is invariant under the LF chiral transformation.
Now substituting $\chi=\chi^{(0)}+\chi^{(m)}$ into the Lagrangian, 
  the first term is decomposed into apparently invariant and 
  (seemingly) non-invariant terms
$$
\psi^\dagger \omega_{\rm EOM}=\psi^\dagger  \left[i\del_+\psi 
-\frac{1}{\sqrt2}\left(\sigma^i\del_i \chi^{(0)}+m_0\chi^{(m)}\right) \right]
+\psi^\dagger \left[-\frac{1}{\sqrt2}
              \left(\sigma^i\del_i \chi^{(m)} +m_0\chi^{(0)}\right) 
              \right]~.
$$
The first term consists of the $m_0$-independent term and quadratically 
  dependent term ${\cal O}(m_0^2)$, while the second term linearly depends 
  on $m_0$.
The ${\cal O}(m_0)$ term changes under the chiral transformation, 
  but due to the relation (\ref{relation}), it eventually vanishes 
  and therefore the Lagrangian is invariant even if there is a mass term.
As a result, we have a conserved Noether current~\cite{Mustaki}
\BQA
&&j_{\rm 5Free}^{\mu}=\bar \Psi \Gmu \gfive \Psi 
   - m_0 \bar \Psi \Gmu\gfive \frac{1}{i\del_-}\gamma^+\psi_+~,\NN
&&\del_\mu j^\mu_{\rm 5Free}=0~, \nonumber
\EQA
which of course reduces to the usual current in the massless limit.

Now let us consider the NJL model.
Decomposition of $\chi$ is straightforward:
\BQA
&&\pmatrix{
 \chi^{(0)}_{1a}\cr
 -\chi^{(0)\dagger}_{2a}
 }
=-\frac{1}{\sqrt2}\int_{-\infty}^{\infty} dy^- G_{ab}(x^-,y^-,x_\perp)
\pmatrix{
 \del_z\psi_{2b}\cr
 \del_z\psi_{1b}^\dagger},\label{chi0}\\
&&\pmatrix{
\chi^{(m)}_{1a}\cr
-\chi^{(m)\dagger}_{2a}
}
=\frac{m_0}{\sqrt2}\int_{-\infty}^{\infty} dy^- G_{ab}(x^-,y^-,x_\perp)
\pmatrix{
\psi_{1b}\cr
\psi_{2b}^\dagger
}.\label{chim}
\EQA
Since the matrix ${\cal A}$, and thus $G_{ab}(x,y)$ is invariant under 
  the transformation (\ref{Chiral_Transf}), it is easy to find that 
  $\chi^{(0)}$ and $\chi^{(m)}$ transform as Eqs.~(\ref{transf0}) and 
  (\ref{transfm}).
Therefore, if $m_0=0$, the LF chiral transformation (\ref{Chiral_Transf}) 
  is equivalent to the usual chiral transformation.
The chiral current and the chiral charge are given by
\BQA
&&j_5^\mu= \bar \Psi \Gmu \gfive \Psi~, \label{Chiral_current}\\
&&Q^{\rm LF}_5=\int_{-\infty}^{\infty} dx^-d^2x_\perp j_5^+ (x)
= \int_{-\infty}^{\infty} dx^-d^2x_\perp \psi^\dagger \sigma^3 \psi~.
\label{charge}
\EQA
How about the massive case?
As we explicitly showed above, the mass term does not prevent 
  chiral symmetry in the free fermion case.
We must bare in mind such possibility even in the NJL model. 
Thus it is worth while to check whether the {\it massive} 
   NJL model is invariant under the LF chiral transformation. 
To see this, it is convenient to treat the Hermite Lagrangian 
\BQ
{\cal L}_{\rm Hermite}=\frac{1}{2}i \psi^\dagger 
\buildrel\leftrightarrow\over{\del}_+\psi 
-\frac{1}{2\sqrt{2}} \left[
  \left( \psi^\dagger \sigma^i\del_i\chi
  +\del_i\chi^\dagger \sigma^i \psi   \right)
  +m_0\left(\psi^\dagger\chi + \chi^\dagger\psi \right) \right] ~.
\EQ
Note that this is equivalent to the free Lagrangian except that 
  $\chi$ is a solution of Eq.~(\ref{FC}). 
Now the apparently non-invariant term is a term linearly depending on $m_0$:
\BQ
-\frac{1}{2\sqrt2}\psi^\dagger 
  \left(\sigma^i\del_i\chi^{(m)}+m_0\chi^{(0)}\right) 
+ {\rm H.c.}~.
\EQ
In the massive free fermion case, we had the same term but 
  it eventually vanished due to Eq.~(\ref{relation}). 
However, in the NJL model, it is evident from Eqs.~(\ref{chi0}) and 
  (\ref{chim}) such relation does not hold because $G$ depends on $x_\perp$.
Therefore we have verified that the massive NJL model is {\it not} 
  invariant under the LF chiral transformation.
If and only if $m_0=0$, the LF chiral transformation is the symmetry 
  of the NJL model and equivalent to the usual chiral transformation. 
This is of course not a surprising result but must be checked explicitly.
Anyway, we do not stick to this problem anymore.

Irrespective of whether we have a mass term or not, we always use the 
  definition for the chiral current Eq.~(\ref{Chiral_current}) which 
  was derived for the massless fermion.
In the massless case, it is of course a conserved current 
  $\del_\mu j^\mu_5=0$, while in the massive case, a usual relation holds
\BQ
\del_\mu j^\mu_5=2m_0 \bar\Psi i\gamma_5 \Psi~, \label{current_div}
\EQ
which is derived by using the Euler-Lagrange equation for the massive fermion.
Equation~(\ref{current_div}) is used 
  when we discuss nonconservation of the chiral charge and 
  the PCAC relation in Sec.~IV.

\subsection{Implication of the fermionic constraint}

So far we treated the fermionic constraint in classical theory 
 and obtained the exact solution (\ref{classical_solution}).
However, this solution does not give a nonzero condensate and 
  the resulting Hamiltonian does not describe the broken phase.
The situation is very similar to the previous analysis of 
  the chiral Yukawa model~\cite{I}.
The chiral Yukawa model in the DLCQ approach has three constraint equations.
We solved them in classical theory but we could not find any way 
  to describe the broken phase with the classical solutions.
What we finally found is that it is very important to treat the constraint 
  equations, especially the zero-mode constraint for a scalar field, 
  nonperturbatively in quantum theory.
This fact is true of our present case.
To obtain a nonzero condensate, we must treat the fermionic constraint 
 as an operator equation and solve it with some nonperturbative method.

To strengthen this, let us briefly overview the procedure 
  in the previous work~\cite{I}. 
In the chiral Yukawa model, we have three dependent variables: 
Two of them are the longitudinal zero-modes of scalar and pseudoscalar fields
  $\sigma_0(x_\perp)=(1/2L)\int_{-L}^L dx^- \sigma(x),\  
  \pi_0(x_\perp)=(1/2L)\int_{-L}^L dx^- \pi(x)$ 
  where $L$ is an extension of the longitudinal direction  $x^-\in [-L,L]$,
  and the rest is the bad component of a fermion $\psi_-(x)$.  
So there are three constraints:
\BQ
   \left( \frac{\mu^2}{\lambda} - \partial_\bot^2 \right)
         \left(\matrix{\sigma_0 \cr \pi_0}\right)
       = \frac{\mu^2}{N} \frac{1}{\sqrt{2}}
           \int_{-L}^L  \frac{dx^-}{2L} \left[ 
           \psi_+^{a \dag} 
           \left(\matrix{-1\cr i\gamma_5}\right)
           \gamma^- \psi^a_- 
         + \psi_-^{a \dag} 
           \left(\matrix{-1\cr i\gamma_5}\right)
           \gamma^+ \psi^a_+ 
           \right]~,
  \label{ZMC}
\EQ
\BQ
  2i \partial_- \psi^a_- = \left( i\gamma^i_\perp\partial_i
             + m_0 +\sigma - i \pi \gamma_5 \right)\gamma^+ \psi^a_+ ~,
  \label{acaa}
\EQ
where $\lambda=g^2N$ in the present notation and 
  $\mu$ is a dimensionless parameter which controls the scalar 
  and pseudoscalar masses.
In the infinitely heavy mass limit, $\mu\rightarrow \infty$, 
 we recover the NJL model.
The procedure of Ref.~\cite{I} is as follows: 
First, we formally solved the fermionic constraint (\ref{acaa}) 
  and substitute the solution into the zero-mode constraints (\ref{ZMC}).
Second, we solved the zero-mode constraints by $1/N$ expansion 
  with a fixed operator ordering and found 
  that the leading order of the scalar zero-mode constraint can be 
  identified with the gap equation. 
Selecting a nontrivial solution of the gap equation, we again 
  substitute it back to the fermionic constraint.
Then we obtain the final expression for the bad component $\psi_-$ 
  in terms of independent variables.
Thus we solved three coupled equations step by step.
On the other hand, we have only one constraint equation.
The procedure in the chiral Yukawa model suggests that 
  we will have to do almost the same procedure 
  {\it at once} when we solve the fermionic constraint. 
Note that just the same as in the chiral Yukawa model, 
  a perturbative solution cannot reach the broken phase 
  even in quantum theory.
Therefore, we naturally expect that solving the fermionic constraint
  (\ref{FC}) in quantum theory using some nonperturbative method is 
  necessary for describing the chiral symmetry breaking~\cite{Itakura}.

\setcounter{equation}{0}

%
%
\section{Solving the fermionic constraint by $1/N$ expansion 
         in quantum theory}

As we discussed above, it is important to solve the fermionic constraint 
  (\ref{FC}) in quantum theory by some nonperturbative method.
Here we solve it with a fixed operator ordering by using the $1/N$ expansion.
For systematic $1/N$ expansion, we introduce the bilocal formulation.
We rewrite the fermionic constraint in terms of bilocal fields and 
  expand it following the Holstein-Primakoff type expansion of the 
  boson expansion method.
We always work with fixed $x^+$.

\subsection{Quantization and the operator ordering}

To solve the constraint in quantum theory, 
  we must first perform the Dirac quantization for constraint systems\footnote{
  In Ref.~\cite{Bentz}, the authors solved the constraint equations for 
  auxiliary fields before canonical quantization was specified and 
  gave a {\it c}-number to the scalar auxiliary field in leading order 
  of $1/N$. Nevertheless, the condensation in the NJL model is a quantum 
  phenomenon and thus this procedure is not justified.}.
After tedious but straightforward calculation of the Dirac brackets, 
  we find a familiar relation for the good component $\psi_\alpha~(\alpha=1,2)$
\BQ
\left\{\psi_\alpha^a(x),\ \psi_\beta^{b\dagger}(y)\right\}_{x^+=y^+}
=\delta_{\alpha\beta}\delta_{ab}\delta(x^--y^-)\delta^{(2)}(x_\perp-y_\perp)~,
  \label{Quantization}
\EQ
and so on.
We introduce the simplest mode expansion at $x^+=0$ 
  as in Ref.~\cite{Leutwyler_meson}:
\BQ
\psi^a_\alpha(x)=
\int_{-\infty}^{\infty}\frac{d^2k_\perp}{2\pi}
\int_0^{\infty}\frac{dk^+}{\sqrt{2\pi}}
\left[ ~
b^a_\alpha(\k)e^{i\k\x}+d^{a\dagger}_\alpha(\k)e^{-i\k\x}
~\right],\label{mode_exp}
\EQ
where $\k\x\equiv -k^+x^-+k_\perp^ix_\perp^i$.
The vacuum is defined by the annihilation operators as
\BQ
b^a_\alpha(\k)\ket{0}=d^a_\alpha(\k)\ket{0}=0~. \label{vac}
\EQ
When we deal with the quantum fermionic constraint, we have to specify 
  the operator ordering.
In many publications discussing the zero-mode constraints, 
  people often choose the Weyl ordering with respect to both constrained 
  and unconstrained variables.
However, in a previous paper~\cite{I}, we discussed that the ideal
  situation was to find a ``consistent'' operator ordering.
For example, let us consider an anticommutator $\{\chi, \psi\}$ in 
  the NJL model.
It can be evaluated in two different ways:
(i) by using the solution $\chi_{\rm sol}=\chi(\psi)$ of the fermionic 
   constraint and the standard quantization rule (\ref{Quantization}), and 
(ii) by calculating the Dirac bracket for $\{\chi, \psi\}$. 
For the case (i), we select a specific operator ordering for 
  the fermionic constraint, and the result depends on the ordering.
For the case (ii), we must also determine the ordering in the r.h.s. of 
  the Dirac bracket $\{\chi, \psi\}_{\rm D}=\cdots$.
These two results must be equivalent to each other.
We have two ambiguity of the operator ordering: those of the constraint 
 equation in (i) and the right hand side of the Dirac bracket in (ii).
``Consistent operator ordering'' should be imposed so that 
  these two quantities be identical.
In other words, we determine the operator ordering of the r.h.s. in the 
  Dirac brackets so that it coincides with the direct evaluation.
In the chiral Yukawa model, we could not check that the ordering 
  we adopted was consistent or not.
Again in the NJL model, this is a very difficult task and 
  we choose a specific operator ordering defined by Eq.~(\ref{FC}).
However, the chiral Gross-Neveu model in 1+1 dimensions allows us to 
 check the consistency of this operator ordering. 
This is briefly shown in Appendix C.

\subsection{Boson expansion method as $1/N$ expansion of bilocal operators}

How can we solve the ``operator equation'' Eq.~(\ref{FC}) by the 
  $1/N$ expansion?
It is generally difficult to count the order ${\cal O}(N^n)$
  of an operator instead of its matrix element.
What is worse, it is physically hard to justify the $1/N$ expansion 
  of the fermionic field itself.
However, as was discussed in Ref.~\cite{BEM_itakura}, 
  there is a powerful method to this problem.
We can perform a systematic $1/N$ expansion of operators 
  if we introduce the bilocal operators and use the boson expansion method.
The boson expansion method is one of the traditional techniques in 
  nonrelativistic many-body problems~\cite{BEM}.
Originally this was invented for describing bosonic excitations in
  non-bosonic systems such as collective excitation in nuclei or spin systems.

Let us rewrite the fermionic constraint (\ref{FC}) in terms of 
  bilocal operators.
We introduce the following ``color'' singlet bilocal operators 
  at equal light-front time 
\BQA
&&\M_{\alpha\beta}(\x, \y)= 
\sum_{a=1}^N \psi^{a \dagger}_\alpha(x^+,\x)\psi^a_\beta(x^+,\y)~,\\
&&\T_{\alpha\beta}(\x, \y) =
\frac{1}{\sqrt{2}}\sum_{a=1}^N \left( \psi^{a \dagger}_\alpha(x^+,\x)
\chi^a_\beta(x^+,\y)
 + \chi^{a \dagger}_\beta(x^+,\y) \psi^a_\alpha(x^+,\x) \right)~,\\
&&\U_{\alpha\beta}(\x, \y) =
\frac{-i}{\sqrt{2}}\sum_{a=1}^N \left(
  \psi^{a \dagger}_\alpha(x^+,\x) \chi^a_\beta(x^+,\y) -
  \chi^{a \dagger}_\beta(x^+,\y)
\psi^a_\alpha(x^+,\x)
   \right)~.
\EQA
We define the Fourier transformation of them as
$$
\M_{\alpha\beta}(\p,\q)=\int_{-\infty}^\infty\frac{d^3\x}{(2\pi)^{3/2}}
\int_{-\infty}^\infty\frac{d^3\y}{(2\pi)^{3/2}}\ 
\M_{\alpha\beta}(\x, \y)\ e^{-i\p\x-i\q\y}~,
$$
and so on. 
Note that this definition allows the longitudinal momenta to take 
  negative values.
Using these bilocal operators, the fermionic constraint (\ref{FC})
  is equivalently rewritten as 
\BQA
i \frac{\partial}{\partial y^-} \T_{\alpha\beta}(\x, \y)
  &=& \frac{1}{2}\left\{ -\partial^y_i 
       \left(
       \sigma^i_{\beta\gamma} \M_{\alpha\gamma}(\x,\y)
     - \sigma^i_{\gamma\beta} \M_{\gamma\alpha}(\y,\x) 
       \right)
     + m_0\Big( \M_{\alpha\beta}(\x,\y)-\M_{\beta\alpha}(\y,\x) 
          \Big) 
                 \right\}                                      \NN
  & &-\frac{g^2}{2}
          \left\{ \M_{\alpha\gamma}(\x,\y) 
                 \Big(\delta_{\gamma\beta} \T(\y,\y)
             + i \sigma^3_{\gamma\beta}~\U(\y,\y) 
                 \Big)  
          \right.                                             \NN
 && \hskip2.7cm  \left.  
                      - \Big( \delta_{\beta\gamma} \T(\y,\y)
                    - i \sigma^3_{\beta\gamma}~\U(\y,\y) 
                        \Big) 
                    \M_{\gamma\alpha}(\y,\x)
                 \right\} ~,                              \label{FC_bilocalT}
\EQA
and 
\BQA
  i^2 \frac{\partial}{\partial y^-} \U_{\alpha\beta}(\x, \y) 
 &=& \frac{1}{2}\left\{ -\partial^y_i 
           \Big(        
              \sigma^i_{\beta\gamma} \M_{\alpha\gamma}(\x,\y) 
            + \sigma^i_{\gamma\beta} \M_{\gamma\alpha}(\y,\x) 
           \Big)
          + m_0\Big( \M_{\alpha\beta}(\x,\y) + \M_{\beta\alpha}(\y,\x) 
               \Big) 
                \right\}                                       \NN
 & & -\frac{g^2}{2}\left\{ \M_{\alpha\gamma}(\x,\y) 
           \Big(
              \delta_{\gamma\beta} \T(\y,\y)
           + i\sigma^3_{\gamma\beta}~\U(\y,\y) 
           \Big)    
                  \right.                                     \NN
 & & \hskip2.7cm  \left.  
            +   \Big(  
                     \delta_{\beta\gamma} \T(\y,\y)
                 - i \sigma^3_{\beta\gamma}~\U(\y,\y) 
                \Big) \M_{\gamma\alpha}(\y,\x)
                  \right\} ~,                            \label{FC_bilocalU}
\EQA
where we have introduced quantities
 $ \T(\x,\y) \equiv \T_{\alpha\alpha}(\x, \y)$ and
 $\U(\x,\y) \equiv (\sigma^3)_{\alpha \beta}~\U_{\alpha\beta}(\x, \y)$
 so that $\bar\Psi\Psi(x)=\T(\x,\x)$ and 
 $\bar\Psi i\gamma_5\Psi(x)=\U(\x,\x)$.
In actual calculation, it is more convenient to treat equations 
  for the operators without spinor structure $\T(\x,\y)$ and $\U(\x,\y)$
  because they form closed equations (see Appendix D).
Once we solve them, we immediately obtain $\T_{\alpha\beta}(\x,\y)$ 
  and $\U_{\alpha\beta}(\x,\y)$ from the above equations.

For systematic $1/N$ expansion of the bilocal fermionic constraints, 
  one must know how to expand $\M_{\alpha\beta}(\p,\q)$.
It is the boson expansion method, especially, the Holstein-Primakoff type 
  expansion for large $N$ theories, that enables us to expand 
  $\M_{\alpha\beta}(\p,\q)$ as {\it operator} quantities:
\BQ
\M_{\alpha\beta}(\p,\q)=N\sum_{n=0}^\infty \left(\frac{1}{\sqrt{N}}\right)^n
\mu_{\alpha\beta}^{(n)}(\p,\q)~.
\EQ
According to the Holstein-Primakoff expansion 
  [Eqs.~(\ref{BEM_-+})$\sim$(\ref{BEM_--})], the first three terms are
  written in terms of bilocal bosonic variable $B(\p,\q)$ as 
\BQA
\mu_{\alpha\beta}^{(0)}(\p,\q)
    &=&\delta_{\alpha\beta}\delta^{(3)}(\p+\q)\theta(p^+)\theta(-q^+)~,
    \label{BEM_0}\\
\mu_{\alpha\beta}^{(1)}(\p,\q)
    &=& B_{\beta\alpha}(\q,\p)\theta(p^+)\theta(q^+)
      + B_{\alpha\beta}^\dagger (-\p,-\q)\theta(-p^+)\theta(-q^+)~,
    \label{BEM_1} \\
\mu_{\alpha\beta}^{(2)}(\p,\q)
    &=& \int[d\k]\sum_\gamma 
         B^\dagger_{\alpha\gamma}(-\p,\k)B_{\beta\gamma}(\q,\k)
         \theta(-p^+)\theta(q^+)\NN
    &-&\int[d\k]\sum_\gamma 
         B^\dagger_{\gamma\beta}(\k, -\q)B_{\gamma\alpha}(\k,\p)
         \theta(p^+)\theta(-q^+)~.\label{BEM_2}
\EQA
where
$$
\int [d\q]=\int_0^\infty dq^+ \int_{-\infty}^\infty d^2q_\perp ~.
$$
Any commutator between $\M_{\alpha\beta}(\p,\q)$'s 
  [such as Eq.~(\ref{algebra})] is correctly reproduced if one uses 
  the following bosonic commutators
\BQA
&&{}\Big[B_{\alpha\beta}(\p_1,\p_2),\ B_{\gamma\delta}^\dagger(\q_1,\q_2)\Big]
=\delta_{\alpha\gamma}\delta_{\beta\delta}
\delta^{(3)}(\p_1-\q_1)\delta^{(3)}(\p_2-\q_2)~,\label{boson_commutator1}\\
&&{}\Big[B_{\alpha\beta}(\p_1,\p_2),\ B_{\gamma\delta}(\q_1,\q_2)\Big]=0 \ \ \
(p_i^+,q_i^+ > 0 )~.\label{boson_commutator2}
\EQA
Note also that the state annihilated by $B(\p,\q)$ is identified with the 
  original Fock vacuum:
\BQ
B(\p,\q)\ket{0}=0~.
\EQ
More detailed discussions about the boson expansion method applied 
  to LF field theories are found in Ref.~\cite{BEM_itakura} and Appendix D
  of the present paper.

\subsection{Solution to the bilocal fermionic constraint}

We are ready to solve the bilocal fermionic constraint 
  using the $1/N$ expansion.
As we commented before, it is convenient to solve the equations for 
  $\T(\p,\q)$ and $\U(\p,\q)$ (see Eqs.~(\ref{FCT}) and (\ref{FCU}) in 
  Appendix~D for their explicit forms).
Once we know $\T(\p,\q)$ and $\U(\p,\q)$, then it is straightforward 
  to obtain $\T_{\alpha\beta}(\p,\q)$ and $\U_{\alpha\beta}(\p,\q)$.

Expanding $\T(\p,\q)$ and $\U(\p,\q)$ as
\BQA
&&\T(\p,\q)=N\sum_{n=0}^\infty\left(\frac{1}{\sqrt{N}}\right)^n 
                t^{(n)}(\p,\q)~,\\
&&\U(\p,\q)=N\sum_{n=0}^\infty\left(\frac{1}{\sqrt{N}}\right)^n 
                u^{(n)}(\p,\q)~,
\EQA
and inserting them into the fermionic constraints, we find for
  the lowest order ${\cal O}(N)$
\BQ
\pmatrix{
t^{(0)}(\p,\q)\cr
u^{(0)}(\p,\q)}
= 
\pmatrix{
m_0\frac{\epsilon(p^+)}{q^+}\delta^{(3)}(\p+\q)\cr
0
}
-g_0^2 \frac{\epsilon(p^+)}{q^+}\int_{-\infty}^\infty
\frac{d^3\k}{(2\pi)^3}
\pmatrix{
   t^{(0)}(\k,\p+\q-\k)\cr
   u^{(0)}(\k,\p+\q-\k)\cr
},\label{LOFC}
\EQ
where $g_0^2=g^2N$. 
Since there are no operators in these equations, $t^{(0)}$ and $u^{(0)}$ 
  are $c$-numbers.  
Nonzero solutions give leading order contribution to 
  $\langle\bar\Psi\Psi\rangle$ and $\langle\bar\Psi i\gfive\Psi\rangle$,
\BQA
&&\bra{0} \bar\Psi\Psi \ket{0}
      = N \int \frac{d^3\p}{(2\pi)^3}f_t(\p) ~+ ~\cdots ~,\\
&&\bra{0} \bar\Psi i \gfive \Psi \ket{0}
      = N \int \frac{d^3\p}{(2\pi)^3}f_u(\p) ~+ ~\cdots~,
\EQA
where 
$t^{(0)}(\p,\q)=f_t(\p)\delta^{(3)}(\p+\q)$ and 
$u^{(0)}(\p,\q)=f_u(\p)\delta^{(3)}(\p+\q)$.
As  the above equation (\ref{LOFC}) with $m_0=0$ is invariant under the 
  chiral rotation, we can always take $u^{(0)}(\p,\q)=0$.
For the massive case, we also take $u^{(0)}(\p,\q)=0$ and  
  $t^{(0)}(\p,\q)\neq 0$.
Now let us introduce a quantity $M$, which corresponds to the 
  dynamical mass of fermion,
\BQ
M=m_0 -g^2 \langle \bar\Psi\Psi\rangle ~.
\EQ
Then, to obtain $t^{(0)}(\p,\q)$ is equivalent to determining $M$, viz.
\BQ
t^{(0)}(\p,\q)=-M\frac{\epsilon(p^+)}{p^+}\delta^{(3)}(\p+\q)~.
\EQ
In terms of $M$, the leading order fermionic constraint 
  (\ref{LOFC}) is rewritten as 
\BQ
\frac{M-m_0}{M}=g_0^2\int \frac{d^3\p}{(2\pi)^3}\frac{\epsilon(p^+)}{p^+}~.
\label{lowest_const_NJL}
\EQ
Physically this equation should be interpreted as a gap equation. 
This is clarified in the next subsection.

Similarly higher order fermionic constraints are solved order by order.
This is because the fermionic constraints for $t^{(n)}(\p,\q)$ and 
  $u^{(n)}(\p,\q)$ are linear equations with respect to the highest order.
The $n=1,2$ solutions are important for giving a nontrivial Hamiltonian 
  and so on.
More details are discussed in Appendix D.

\subsection{Gap equation}

Now let us discuss the physics meaning of Eq.~(\ref{lowest_const_NJL}).
As we mentioned above, this equation should be regarded as a gap equation
  for chiral condensate. 
In several previous studies of ours, we have seen essentially the same 
  kind of equations~\cite{ItakuraMaedan,Itakura,I}. 
Indeed, in Ref.~\cite{Itakura} it was pointed out that 
  Eq.~(\ref{lowest_const_NJL}) itself is the gap equation.
Also in the chiral Yukawa model~\cite{I},  the zero-mode 
  constraint for the scalar field reduced to the above equation and 
  was interpreted as a gap equation.
Since this identification is an indispensable step for our framework,
  let us again explain it within the NJL model.

First of all, consider a naive massless limit $m_0\rightarrow 0$ of 
  Eq.~(\ref{lowest_const_NJL}):
$$
M\left(1-g_0^2\int \frac{d^3\p}{(2\pi)^3}\frac{\epsilon(p^+)}{p^+}\right)=0~.
$$
Thus we find two possibilities: the first is $M=0$ and the second is 
\BQ
1-g_0^2\int \frac{d^3\p}{(2\pi)^3}\frac{\epsilon(p^+)}{p^+}=0
\label{gap_massless}
\EQ but $M$ 
  is arbitrary.
Of course the first case is not interesting because it 
  corresponds to the symmetric phase. 
On the other hand, the second case with nonzero $M$ does not immediately 
  mean the existence of broken phase.
Since the equation (\ref{gap_massless}) is independent of $M$ as it is,
  the dynamical mass $M$ is left undetermined,  which is not a physically 
  acceptable situation.
However, this observation is not correct 
   because the divergent integral in Eq.~(\ref{gap_massless}) is not 
   regularized. 
Indeed, we can identify  Eq.~(\ref{gap_massless}) with the gap equation  
  only after we carefully treat the infrared (IR) divergence.

To see this, let us put an IR cutoff.
First consider the same cutoff schemes as in the equal-time formulation, 
   such as the covariant four-momentum cutoff.
We can easily translate it into a cutoff on the light-cone momentum 
  $k^+$ and $k_\perp^i$ and obtain the correct gap equation.
Indeed, in Ref.~\cite{Heinzl}, a noncovariant (rotationally
   invariant) three-momentum cutoff was performed to obtain the known result.
But such a cutoff is artificial as a light-front theory, 
   and we here adopt another cutoff scheme, {\it the parity invariant cutoff}.
Usually, it is natural and desirable to choose a cutoff so as to preserve 
   symmetry of a system as much as possible.
For the LF coordinates $x^\pm$ and $x^i_\perp$, it would be 
   natural to consider parity transformation 
   ($x^+\leftrightarrow x^-,\ x^i_\perp\rightarrow -x^i_\perp$) 
   and two-dimensional rotation in the transverse plane.
In the usual canonical formulation where $x^+$ is treated 
   separately, the parity invariance is not manifest.
However, we find it useful for obtaining the gap equation.
In momentum space, the parity transformation
   is exchange of $k^+$ and $k^-$ and replacement 
   $k_\perp^i\rightarrow - k_\perp^i$.
Therefore the parity invariant cutoff is given by $k^\pm<\Lambda$ 
  and $k_\perp^2<\Lambda'^2$.
Using the dispersion relation $2k^+k^--k_\perp^2=M^2$,
   we find that the parity invariant regularization 
   inevitably relates the ultraviolet (UV) and IR cutoffs: 
\BQ
   \frac{k_\perp^2+M^2}{2\Lambda}<k^+<\Lambda~.\label{PIC}
\EQ
This also implies the planar rotational invariance
    $k_\perp^2<2\Lambda^2-M^2=\Lambda'^2$.
What is important here is the use of constituent mass $M$ in the 
   dispersion relations. 
Physically it corresponds to imposing {\it self-consistency conditions}.
Since the IR cutoff includes $M$, the r.h.s. of 
   Eq.~(\ref{lowest_const_NJL}) has nontrivial dependence on $M$:
\BQ
\frac{M-m_0}{M}=\frac{g_0^2\Lambda^2}{4\pi^2}\left\{
2-\frac{M^2}{\Lambda^2}\left( 1 + {\rm ln}\frac{2\Lambda^2}{M^2}  \right)
\right\}~.\label{gap_NJL}
\EQ
This is the gap equation and is equivalent to that of the previous 
  result in the chiral Yukawa model~\cite{I}.
It has a nonzero solution $M\neq 0$ even in the $m_0\rightarrow 0$ limit.
The somewhat unfamiliar equation (\ref{gap_NJL}) of the NJL model 
   exhibits the same property 
   as the standard gap equations of the equal-time quantization.
For example, when $m_0=0$, there is a critical coupling 
   $g_{\rm cr}^2=2\pi^2/\Lambda^2$, above which $M\neq 0$.

The essential and inevitable 
   step to obtain the gap equation is the inclusion of mass information
   as the regularization rather than the fact that the UV and IR
   cutoffs are related to each other.
If we regulate the divergent integral without mass information 
   ( e.g., introducing the UV and IR cutoffs independently),
   we cannot reproduce the gap equation.
The loss of mass information is closely related to 
   the fundamental problem of the LF formalism~\cite{Naka-Yama}, 
   and the parity invariant 
   regularization can be considered as one of the prescriptions for it.
Reference~\cite{Naka-Yama} discussed within scalar theory that {\it the 
   light-front quantization gives a mass-independent two-point function} 
   (at equal LF time), which contradicts the result from general
   arguments concerning the spectral representation.
We have been encountered with the same problem in Eq.~(\ref{lowest_const_NJL})
   because the integral is regarded as a naive estimation of 
   $\langle\bar\Psi\Psi\rangle/M$ by using fermion with mass $M$.   
And also the origin of mass-independent result can be traced back to 
   the mode expansion (\ref{mode_exp}).
Even if we include the wave function for free fermion field, we do not have 
   any mass-dependence on the mode expansion~\cite{Heinzl_review}.

Let us give a brief comment on the chiral Gross-Neveu model.
Of course the important difference of the 1+1 dimensional case 
  is the renormalizability.
Ignoring the transverse directions in the above calculation, we easily find 
  the gap equation  $(M-m_0)/M=g^2_0/(2\pi) \ln (2\Lambda^2/M^2)$ 
  where the parity-invariant cutoff $M^2/2\Lambda <k^+<\Lambda$ was used.
Though it explicitly depends on the cutoff $\Lambda$ and is  divergent as
  $\Lambda\rightarrow \infty$, we can remove the divergence by 
  coupling constant renormalization \cite{Itakura}.

\subsection{Hamiltonian}
Having the solution to the bilocal fermionic constraint, we can rewrite 
   the fermion bilinear operators in terms of the bilocal bosons.
Of special importance is the (Hermitian) Hamiltonian, which is easily 
  expressed by $\T_{\alpha\beta}(\p,\q)$ and $\U_{\alpha\beta}(\p,\q)$ 
  as follows
\BQA
H=P^-&=& \frac{1}{2\sqrt{2}}\int d^3x  \left[
  \left( \psi^\dagger \sigma^i\del_i\chi
  +\del_i\chi^\dagger \sigma^i \psi   \right)
  +m_0 \left(\psi^\dagger\chi + \chi^\dagger\psi \right) \right]   \NN
&=& \frac{1}{4} \int d^3\p~d^3\q \delta^{(3)}(\p+\q)~
    iq_\perp^i \sigma^i_{\alpha\beta}
   \left[ \Big( \T_{\alpha\beta}(\p,\q)+\T_{\beta\alpha}(\q,\p) \Big)
     +i \Big(\U_{\alpha\beta} (\p,\q) - \U_{\beta\alpha}(\p,\q) \Big)
  \right]  \nonumber  \\
 && + \frac{m_0}{2} \int d^3\p~d^3\q \ \T (\p,\q)~\delta^{(3)}(\p+\q)~.
\label{Ham_NJL}
\EQA
Apparently this Hermitian version of the Hamiltonian seems equivalent to
   the free Hamiltonian, but the information of interaction enters
   through the bad component $\chi$.
We find the $1/N$ expansion of the Hamiltonian 
\BQ
H=N\sum_{n=0}^\infty \left(\frac{1}{\sqrt{N}}\right)^n h^{(n)}~,
\EQ
by substituting the solutions of the fermionic constraints into 
  the Hamiltonian.
The zeroth order contribution is just a divergent constant and we discard it.
The first  order is strictly zero. 
Nontrivial contribution comes from the order ${\cal O}(N^0)$
\BQA
h^{(2)}&=&\int [d\p][d\q]  \left( \frac{p^2_\perp +M^2}{2p^+} 
        + \frac{q^2_\perp +M^2}{2q^+} \right) 
          B^\dag_{\alpha \beta}(\p,\q ) B_{\alpha \beta}(\p,\q)  \NN
   && + \frac{g_0^2}{(2\pi)^3} \int[d\p][d\q][d\k][d\itl]
        \delta^{(3)}(\p+\q-\k-\itl )\alpha(p^++q^+)                    \NN
   &&\quad \times \left[  \sum {}^{\alpha\beta}_{\gamma\delta}(\p,\q;\k,\itl)
          - \prod {}^{\alpha\beta}_{\gamma\delta}(\p,\q;\k,\itl)\right] 
          B^\dag_{\alpha \beta}(\p,\q ) B_{\gamma\delta}(\k,\itl )  \NN
 && +\ c\rm{~number}~, \label{Ham_NJL2}
\EQA
where $\alpha(p^++q^+)$ is defined by Eq.~(\ref{alpha}) and 
  ``kernels'' of the interaction terms are
\BQA
&&\sum {}^{\alpha\beta}_{\gamma\delta}(\p,\q;\k,\itl)
   \equiv \Big[\SS (-\p)-\SS (-\q)\Big]_{\alpha\beta}
    \Big[\SS (\k)- \SS (\itl)\Big]_{\delta\gamma}  ~, \label{kernel_sigma}\\
&&\prod {}^{\alpha\beta}_{\gamma\delta}(\p,\q;\k,\itl)
   \equiv \Big[\PP (-\p) - \PP (\q) \Big]_{ \alpha \beta} 
    \Big[\PP (-\k) - \PP (\itl) \Big]_{\delta \gamma} ~,\label{kernel_pi}
\EQA
with
\BQ
\SS_{\alpha\beta}(\p) 
    = \left(\frac{ip^i \sigma^i-M}{2p^+}\right)_{\alpha\beta}~,\quad
\PP_{\alpha\beta}(\p) 
    = \left( \frac{ip^i \sigma^i -M}{2p^+} \sigma^3\right)_{\alpha\beta}~.
\EQ
As is evident from the explicit forms of the kernels (\ref{kernel_sigma})
  and (\ref{kernel_pi}), they originate from the scalar interaction 
  $(\bar\Psi\Psi)^2$ and the pseudoscalar one $(\bar\Psi i\gfive \Psi)^2$,
  respectively.
If we substitute a nontrivial (trivial) solution of the gap equation 
  (\ref{gap_NJL}) into the above Hamiltonian, then it governs dynamics 
  of the broken (symmetric) phase.
The first term of $h^{(2)}$ corresponds to a free part with mass $M$
  and the second term to an interaction part.
In the broken phase, $M$ is the dynamical mass and the Hamiltonian 
  suggests a constituent picture.

As we mentioned before, the Hermite Hamiltonian of the chiral Gross-Neveu 
  model has only an $m_0$-dependent term. 
Neglecting the transverse coordinates in Eq.~(\ref{Ham_NJL}), we have
$$
P^-_{\rm GN}= \frac{m_0}{2\sqrt{2}}\int dx^-  
     \left(\psi^\dagger\chi + \chi^\dagger\psi \right)  ~.
$$
Furthermore, the classical solution for the bad spinor component $\chi$ 
  is proportional to $m_0$.
Therefore naive $m_0\rightarrow 0$ limit gives just a zero Hamiltonian.
However, if we solve the gap equation and substitute the nontrivial 
  solution into the Hamiltonian, the resulting Hamiltonian turns 
  out to be  proportional to $M^2$ and survives even in the chiral limit.
This is easily seen from the Hamiltonian of the NJL model (\ref{Ham_NJL2}).
The (constituent) mass term in Eq.~(\ref{Ham_NJL2}) comes from the bare mass 
  term, whose factor $m_0$ cancels with a factor $M^2/m_0$  
  in the second order solution $\int d^3\p~t^{(2)}(\p,-\p)$.
Of course this is not reached if we set $m_0=0$ from the beginning.
Therefore inclusion of the bare mass term is necessary to obtain 
  a correct (constituent) mass term of the Hamiltonian.

\setcounter{equation}{0}
%
%
\section{Physics in the broken phase}

By solving the fermionic constraint, we acquired necessary ingredients 
  for discussing physics consequences of the chiral symmetry breaking.
Basic quantities such as $\bar\Psi \Psi$,  $\bar\Psi i\gfive \Psi$, and 
  the null-plane chiral charge (\ref{charge}) are expressed in terms of the 
  bilocal bosons $B_{\alpha \beta}(\p,\q)$ and $B^\dag_{\alpha \beta}(\p,\q)$
  as
\BQA
&&\bar\Psi \Psi(x) = \T(\x,\x)= \frac{N}{g_0^2}(m_0-M)
  + \sqrt{N}\int\frac{d^3\p d^3\q}{(2\pi)^3}~t^{(1)}(\p,\q)~e^{i(\p+\q)\x} 
  \ + \ {\cal O}(N^0) ~,\label{T_op}\\
&&\bar\Psi i\gfive \Psi(x)= \U(\x,\x)=
    \sqrt{N}\int\frac{d^3\p d^3\q}{(2\pi)^3}~u^{(1)}(\p,\q)~e^{i(\p+\q)\x} 
  \ +\  {\cal O}(N^0) ~,\label{U_op}\\
&&Q_5^{\rm LF} = \int d^3\p~\sigma^3_{\alpha \beta}
 : \M_{\alpha\beta}(\p,-\p) : ~= \int d^3\p~\sigma^3_{\alpha \beta} 
                 ~\mu^{(2)}_{\alpha \beta}(\p,-\p)\  +\  {\cal O}(N^{1/2})~,
\label{charge_op}
\EQA
where  $t^{(1)}(\p,\q)$ and $u^{(1)}(\p,\q)$ are given in Appendix D.  
Now that these are given as functions of the bilocal bosons at the 
  operator level, all the calculation is done with the commutators 
  (\ref{boson_commutator1}) and (\ref{boson_commutator2}).

\subsection{Chiral transformation and nonconservation of 
            chiral charge}

Why could we obtain a nonzero fermion condensate? 
To understand this, let us rewrite the fermionic constraint (\ref{FC}) as
$$
i\del_-\chi_a
 =\frac{1}{\sqrt{2}}\left(-\sigma^i\del_i+m_0\right)\psi_a
   -\frac{g^2}{\sqrt2}
   \left( \psi_a \T(\x,\x) + i\sigma^3\psi_a \U(\x,\x) \right)~,
$$
and substitute Eqs.~(\ref{T_op}) and (\ref{U_op}) 
   into $\T(\x,\x)$ and $\U(\x,\x)$, respectively\footnote{In the leading 
   order, this procedure corresponds to the mean-field
   approximation done by Heinzl et al. \cite{Heinzl}. 
   They solved the fermionic constraint by simply linearizing the interaction 
   parts as $-g^2/\sqrt2 \psi_a \langle \T(\x,\x)\rangle$.
   By evaluating the vacuum expectation value $\langle \T(\x,\x)\rangle$ 
   self-consistently with the dynamical fermion mass 
   $M=m_0-g^2\langle \T(\x,\x)\rangle$, they obtained the gap equation. 
   If one uses the parity-invariant cutoff, the result coincides with ours.}.
Then the leading order equation turns out to be equivalent to the
   constraint equation for a free fermion with mass $M$,
$$
i\del_-\chi_a =\frac{1}{\sqrt{2}}\left(-\sigma^i\del_i+M\right)\psi_a ~.
$$
Also at the same order, the equation of motion for the good component 
  $\psi$ says that the fermion acquires a mass $M$.
This means that the operator structure of the bad spinor $\chi$ 
  changes depending on which solutions of the gap equation 
  (\ref{lowest_const_NJL}) is selected.
For massive fermion, the fermion condensate $\langle \bar\Psi\Psi\rangle$ 
  is no longer zero even if the vacuum is trivial.
One can find an analogy between the chiral Yukawa model and the NJL model 
  because in the chiral Yukawa model, operator structure of the longitudinal 
  zero modes and subsequently of the bad spinor component changes depending on
  the phases.

One thing to be noted is the peculiarity of the mode expansion 
  (\ref{mode_exp}).
It is evident that the mode expansion has no mass dependence in it.
This caused the problem of identifying the lowest fermionic constraint with 
  the gap equation. 
We had to supply mass information properly when we regularize the IR 
  divergence.
On the other hand, such independence of mass, in turn, implies that 
  our mode expansion allows fermions with {\it any value of mass}.
In other words, the LF vacuum does not distinguish mass of the fermion.
Therefore we can regard the vacuum for massless fermion as that 
  for massive one.
Mass of the fields is determined by the Hamiltonian.
This is the reason why we can live with the trivial vacuum while having 
  a nonzero fermion condensate.
This fact is not a limited phenomenon for our specific mode expansion
   but a common one for light-front field theories.
Indeed, even if we expand a fermion field with free spinor wave functions, 
  $u(p)$ and $v(p)$, we have no mass dependence~\cite{Heinzl_review}.

The fact that the operator structure changes depending on the phases,
  also resolves a seeming contradiction between 
  the triviality of the null-plane chiral charge 
  and the nonzero chiral condensate $\bra{0}\bar\Psi\Psi\ket{0}\neq 0$.
In general, it is known that a null-plane charge always annihilates 
  the vacuum irrespective of the presence of symmetry.
This can be checked explicitly by the expression (\ref{charge_op}), viz.
\BQ
Q_5^{\rm LF}\ket{0}=0~.
\EQ
However, the triviality of $Q_5^{\rm LF}$ in the presence of 
  the chiral condensate immediately leads to a 
  contradiction if an equation $[Q_5^{\rm LF}, \bar\Psi i\gfive\Psi]
  =-2i\bar\Psi\Psi$  could hold in the broken phase.
In the previous analysis of the chiral Yukawa model~\cite{I}, 
  we were faced with exactly the same problem and resolved it 
  by recognizing that in the broken phase the chiral transformation 
  of {\it dependent} variables are different from the usual one 
  simply because their operator structure changes.
This is of course true of the NJL model.
First of all, as we saw above, if we select the nontrivial solution 
  of the gap equation, the fermion is no longer a massless fermion 
  even in the chiral limit. 
Secondly, we can explicitly show that 
  the usual transformation law $[Q_5^{\rm LF}, \bar\Psi i\gamma_5\Psi ] 
  = - 2i \bar\Psi\Psi$ holds only in the symmetric phase ($M=0$).
In the broken phase a simple calculation (up to ${\cal O}(N^{1/2})$) 
  leads to 
\begin{eqnarray}
 &&\left[ Q_5^{\rm LF}, \bar \Psi i \gamma_5 \Psi (x) \right] \NN
 &&\quad = - 2i \bar \Psi\Psi(x) + 2i \frac{N}{g_0^2}(m_0-M)
     +2i\sqrt{N} M\int \frac{d \p d \q}{(2\pi)^3}\ e^{i(\p+\q)\x}\NN
 &&\qquad \times \left[ \frac{\mu^{(1)}_{\alpha \alpha}(\p,\q)}{q^+} 
    -g_0^2 \frac{\epsilon(p^+)}{q^+}\alpha(p^++q^+) 
              \int \frac{d^3 \k }{(2\pi)^3} 
              \frac{\mu^{(1)}_{\alpha \alpha}(\k,\p+\q-\k)}{p^++q^+-k^+} 
     \right] + {\cal O}(N^0)~.
\end{eqnarray}
Even if we take the chiral limit $m_0\rightarrow 0$, the extra term  
  proportional to $M$ survives nonzero.
This also implies that if we put $M=0$, the usual relation holds.
The unusual chiral transformation, however, is consistent with the 
   triviality of $Q_5^{\rm LF}$ because 
   $\bra{0}[Q_5^{\rm LF},\bar\Psi i\gfive\Psi]\ket{0}=0$.

A similar situation occurs for the Hamiltonian.
Nonconservation of the null-plane chiral charge has been pointed out 
  by several people as a characteristic feature of the chiral symmetry 
  breaking on the LF~\cite{nonconserv,Tsujimaru-Yamawaki}. 
They discussed it under the assumption of the PCAC relation, but we can check 
  it explicitly by using the broken Hamiltonian.
After lengthy calculation, we find the commutator $[Q_5^{\rm LF}, H]$ is 
 really nonzero and again proportional to the dynamical mass $M$:
\BQA
\left[Q_5^{\rm LF} , H \right]
   &=& M\frac{g_0^2}{16\pi^3 i}\int[d\p][d\q][d\k][d\itl]
       \delta^{(3)}(\k+\itl -\p-\q)\alpha(p^++q^+)
       B^\dagger_{\alpha\beta}(\p,\q) B_{\gamma\delta}(\k,\itl )  \NN
   & & \times \left[
                \left( \frac{p^i_\perp}{p^+}- \frac{q^i_\perp}{q^+} \right)
                      \sigma^i_{\alpha\beta }
                \left(\frac{1}{k^+}+\frac{1}{l^+}\right)
                      \sigma^3_{\delta \gamma}
               -\left(\frac{1}{p^+}-\frac{1}{q^+}\right)
                      \delta_{\alpha\beta }
                \left(\frac{k^i_\perp}{k^+}- \frac{l^i_\perp}{l^+}  \right) 
                     (\sigma^i \sigma^3)_{\delta \gamma}
               \right.                                            \NN
   & &  \quad \left.
                -\left(\frac{p^i_\perp}{p^+}-\frac{q^i_\perp}{q^+}\right)
                      (\sigma^i \sigma^3)_{\alpha\beta }
                 \left( -\frac{1}{k^+}+\frac{1}{l^+} \right)
                     \delta_{\delta\gamma}
                +\left( \frac{1}{p^+} + \frac{1}{q^+} \right)
                    \sigma^3_{\alpha \beta}
                 \left( \frac{k^i_\perp}{k^+}- \frac{l^i_\perp}{l^+}\right) 
                    \sigma^i_{\delta\gamma}
               \right]                                            \NN
   &&         +~{\cal O}(N^{-\frac{1}{2}} )~.
\label{nonconserv_H}
\EQA
Therefore, the LF chiral charge is not conserved even in the chiral limit.
In our framework it would be more understandable to mention that {\it the 
  Hamiltonian is not invariant under the LF chiral transformation in the 
  broken phase}.
The broken phase Hamiltonian (\ref{Ham_NJL2}) has three terms: 
  $M$-independent, linearly dependent and quadratically dependent terms.
The quadratically dependent term, as well as the $M$-independent one,
  does not break the LF chiral symmetry.
It is the term proportional to the dynamical fermion mass $M$
  which breaks the LF chiral symmetry.
And also, since Eq.~(\ref{nonconserv_H}) is proportional to $g^2_0$, 
  the symmetry breaking term purely comes from the interaction\footnote{For 
  a massive free fermion, we have $[Q_5^{\rm LF}, H]=0$.}.

This result should be consistent with the current divergence relation
 Eq.~(\ref{current_div}).
Integrating it over the space, we have
\BQ
  \partial_+ Q_5^{\rm LF} = \frac{1}{i} \left[ Q_5^{\rm LF} , H \right] 
                          = 2m_0 \int dx^- d^2 x_\perp 
                                 \bar\Psi i \gamma_5 \Psi ~.
  \label{nonconserv}
\EQ
Therefore if the LF chiral charge is not conserved in the chiral limit, 
  the r.h.s. must show a singular behavior 
\BQ
\int dx^- d^2 x_\perp \bar \Psi i \gamma_5 \Psi \propto \frac{1}{m_0}~.
\label{singular}
\EQ
This can be verified directly by using the solution of the fermionic 
  constraint.
Indeed we find that  $\int dx^- d^2 x_\perp\bar\Psi i\gamma_5\Psi=
  \int d\p~u^{(2)}(\p,-\p)$  is  proportional to $M/m_0$ and 
  gives exactly the same result as Eq.~(\ref{nonconserv_H}).
Importance of such singular behavior for making the Nambu-Goldstone boson 
  meaningful was stressed by Tsujimaru et al. in scalar 
  theories~\cite{Tsujimaru-Yamawaki}.
Assuming the PCAC relation, they showed that the zero mode of the 
  Nambu-Goldstone boson has a singularity of $m_{\rm NG}^{-1}$ where 
  $m_{\rm NG}$ is an explicit symmetry-breaking mass.
Our result (\ref{singular}) is consistent with theirs because 
  the operator $\bar\Psi i\gfive \Psi$ is directly related to
  the Nambu-Goldstone boson.
Later, we will prove that the PCAC relation is derived from the current 
  divergence relation (\ref{current_div}).


\subsection{LF bound-state equations for mesons and their solutions}

\subsubsection{Single bosonic state as a fermion-antifermion state}

In our formulation with the boson expansion method, 
  any bosonic excited state is described by the Fock states 
  of the bilocal bosons constructed on the trivial vacuum:
\BQ
\prod_i B^\dag_{\alpha_i\beta_i}(\p_i,\q_i)\ket{0}~.
\EQ
Since the Hamiltonian (\ref{Ham_NJL2}) is quadratic with respect to the 
  bilocal bosons, the first excited state is given by a single bosonic state.
In fermionic degrees of freedom, the one boson state corresponds to 
  the leading contribution (of $1/N$ expansion) of a fermion-antifermion  
  state.
To see this, let us write a mesonic state only with a ``color'' singlet 
  fermion-antifermion Fock component:
\BQ
\ket{~{\rm meson}; P^+,P_\perp}=\frac{1}{\sqrt{N}}\int_0^{P^+}dk^+
  \int_{-\infty}^{\infty}d^2k_\perp \ 
 \Phi^{\alpha\beta}(\k) 
 b_\alpha^{a\dag}(\k)d_\beta^{a\dag}(\itP-\k)\ket{0}~,
\EQ
where the LC wavefunction $\Phi^{\alpha\beta}(\k)$ is normalized 
 so as to satisfy the condition
\BQ
\bra{~{\rm meson}; \itP~ }
   ~{\rm meson};\mbox{\boldmath $Q$~}\rangle 
  = (2\pi)^32P^+\delta^{(3)}(\itP-\mbox{\boldmath $Q$})~,
  \label{norm}
\EQ
or equivalently,
\BQ
\int_0^1 dx\int\frac{d^2k_\perp}{16\pi^3}\sum_{\alpha\beta}
\vert\Phi^{\alpha\beta}(\k)\vert^2=1.
\EQ
According to the Holstein-Primakoff type expansion (\ref{BEM_--}), 
  the fermion-antifermion operator $b_\alpha^{\dag}d_\beta^{\dag}$ 
  can be equivalently rewritten as
\BQA
b_\alpha^{a\dag}(\k)d_\beta^{a\dag}(\itP-\k)
   &=& :\M^{--}_{\alpha\beta}(-\k,-\itP+\k):   \NN
   &=& \sqrt{N}B^\dag_{\alpha\beta}(\k,\itP-\k) \NN
   & &-\frac{1}{2\sqrt{N}}\int [d\q][d\q']B^\dag_{\gamma\beta}
                             (\q,\itP-\k)
           B^\dag_{\alpha\delta}(\k,\q')B_{\gamma\delta}(\q,\q')+ \cdots ~.
\EQA
Therefore at the leading order of $1/N$ expansion, 
   the mesonic state is described as a single (bilocal) boson state,
\BQ
\ket{~{\rm meson}; P^+,P_\perp}= \int_0^{P^+}dk^+\int_{-\infty}^\infty 
 d^2k_\perp \ 
 \Phi^{\alpha\beta}(\k)B^\dag_{\alpha\beta}(\k,\itP-\k)\ket{0} 
 + {\cal O}(N^{-1/2})~.
\EQ
Besides this, it is evident from the normalization  condition (\ref{norm}),
  a local operator $a^\dag (\itP) =\int d^3k
 \Phi^{\alpha\beta}(\k)B^\dag_{\alpha\beta}(\k,\itP-\k)$
 satisfies the usual bosonic commutators.
 
The LC wavefunction $\Phi^{\alpha\beta}(\k)$ and the mass of a meson
 $M_{\rm meson}$ is determined by solving the light-front eigen-value 
 equation:
\BQ
h^{(2)}\ket{~{\rm meson}; P^+,P_\perp=0}
=\frac{M_{\rm meson}^2}{2P^+}\ket{~{\rm meson}; P^+,P_\perp=0}~, \label{LCBSE}
\EQ
where we set $P^i_\perp=0$, for simplicity.

\subsubsection{Scalar and Pseudoscalar mesons}

In the leading order of $1/N$ expansion, the Hamiltonian has only quadratic 
  terms  of the bosonic operators.
Therefore, diagonalization of the Hamiltonian, or equivalently, 
  solving the light-cone bound state equation (\ref{LCBSE})  
  is straightforward.
First of all, if one notices the orthogonal property 
$\left(\PP(-\k)-\PP(\itl)\right)_{\alpha\beta}
 \left(\SS(\k)-\SS(\itl)\right)_{\beta\alpha}=0
$
where $\k=(x P^+ , k^i_\perp )$ and $\itl=\itP-\k=
 ((1-x)P^+,-k^i_\perp)$, one can easily find 
  the spinor structure for  scalar ($\sigma$) and pseudoscalar ($\pi$) 
  states should be 
\BQA
&&\ket{~\pi~; P^+ , P_\perp =0 } = P^+ \int_{0}^{1} dx \int d^2 k_\perp 
  ~ \phi_\pi (x, k^i _\perp )
 \left\{ \left(i k^i_\perp \sigma^i +M\right)\sigma^3 \right\}_{\alpha \beta}
   B^{\dag}_{\alpha\beta}(\k,\itl) \ket{0}~,\label{pionic_state}\\
&&\ket{~\sigma~; P^+ , P_\perp =0 } = P^+ \int_{0}^{1} dx \int d^2 k_\perp 
  ~ \phi_\sigma (x, k^i _\perp )
   \left\{ i k^i_\perp \sigma^i + (1-2x) M \right\}_{\alpha \beta}
   B^{\dag}_{\alpha\beta}(\k,\itl) \ket{0}~.
\EQA
These two states are orthogonal to each other.
Somewhat nonstandard spinor structure of the mesonic states is due to our 
  specific choice of the mode expansion Eq.~(\ref{mode_exp}) 
  and the representation for $\gamma$ matrices Eq.~(\ref{two-compo-gamma}).
For example, if one rewrites the pseudoscalar field $\bar\Psi i\gfive \Psi$
  in terms of the bilocal bosons, one finds the same spinor structure as 
  that of Eq.~(\ref{pionic_state}).  
Note also that $\left\{\gfive(\gamma_\perp^i k_\perp^i+M)\right\}_{\alpha\beta}= 
  \left\{ \left(i k^i_\perp \sigma^i +M\right)\sigma^3 \right\}_{\alpha \beta}$
  for $\alpha,\beta=1,2$ in our two-component representation for the $\gamma$ 
  matrices.
  
Spinor independent parts of the LC wavefunctions 
  $\phi_{\pi,\sigma}(x, k^i _\perp)$ are given as solutions of the 
  following integral equations:
\BQA
m_\pi^2 \phi_\pi (x, k_\perp^i)
&=& \frac{k_\perp^2+M^2}{x(1-x)} \phi_\pi (x , k_\perp^i)         
 -  \frac{g_0^2 \alpha}{(2\pi)^3}\frac{1}{x(1-x)}
    \int_0^1 dy  \int d^2 \ell_\perp 
      \frac{\ell_\perp^2 + M^2}{y(1-y)}
    \phi_\pi (y , \ell_\perp^i) ~, \label{LFBSE_pion}\\
m_\sigma^2 \phi_\sigma (x,k_\perp^i)&=& \frac{k_\perp^2+M^2}{x(1-x)}
      \phi_\sigma(x , k_\perp^i)                       
   - \frac{g_0^2\alpha}{(2\pi)^3}\frac{1}{x(1-x)}
   \int_0^1 dy \int d^2 \ell_\perp 
   \frac{ \ell_\perp^2 + (1-2y)^2 M^2 }{y(1-y)}
   \phi_\sigma (y , \ell_\perp^i) ~.\label{LFBSE_sigma}
\EQA
Here the factor $\alpha=\alpha(P^+)$ defined by Eq.~(\ref{alpha}) is 
  given as a result of the gap equation,
\BQ
\alpha=\left( \frac {m_0}{M}+\frac{2 g_0^2}{(2\pi)^3} \int d^2 q_\perp
      \int_0^1\frac{dx}{x}\right)^{-1}~.\label{alpha_br}
\EQ
Since these integral equations are separable ones, 
  solutions are easily found 
\BQA
&&\phi_\pi (x, k_\perp^i)
   =-C_\pi \frac{g_0^2}{(2\pi)^3}\frac{M}{m_0}
    \frac{1}{x(1-x)-(k_\perp^2 + M^2) /m_\pi^2}~,\label{solpb}\\
&&\phi_\sigma (x, k_\perp^i)
   =-C_\sigma \frac{g_0^2}{(2\pi)^3}\frac{M}{m_0} 
    \left(\frac{m_\sigma^2 - 4M^2 }{m_\sigma^2} \right)
    \frac{1}{x(1-x)-(k_\perp^2 + M^2)/m_\sigma^2}~,\label{solsb}
\EQA
where $C_\pi$ and $C_\sigma$ are constants 
  $C_{\pi,\sigma}=\int_0^1dx\int d^2k_\perp \phi_{\pi,\sigma}(x,k_\perp^i)$.

Equations for $m_\pi$ and $m_\sigma$ are derived from 
  the normalization condition for the LC wavefunctions, viz.
\BQA
&&1= g_0^2\frac{M}{m_0}\int_{0}^{1} dx \int\frac{d^2 k_\perp}{(2\pi)^3}
     \frac{m_\pi^2}{ k_\perp^2 + M^2 - m_\pi^2 x(1-x) }~,\\
&&1= g_0^2 \frac{M}{m_0} \int_{0}^{1} dx \int \frac{d^2 k_\perp}{(2\pi)^3}
     \frac{m_\sigma^2 - 4M^2}{ k_\perp^2 + M^2 - m_\sigma^2 x(1-x) }~.
\EQA
These are exactly equivalent to the corresponding equations 
  in the previous work on the chiral Yukawa model~\cite{I},
  where we obtained them by calculating pole masses 
  of the scalar and pseudoscalar bosons.
If one uses the same cutoff scheme (extended parity-invariant cutoff) 
  as in Ref.~\cite{I}:
\BQ
\frac{k_\perp^2+M^2}{x}+\frac{k_\perp^2+M^2}{1-x} < 2\Lambda^2,
\EQ
the pseudoscalar mass for small bare mass $m_0$ is 
\BQ
 m_\pi^2 = \frac{m_0 N}{g_0^2 M}Z_\pi + {\cal O}(m_0^2)~, \label{pion_mass}
\EQ
where a cutoff dependent factor  
\BQ
Z_\pi= \frac1N 
 \left[ \frac{1}{8\pi^2} \ln
   \left(\frac{1+\sqrt{1-2M^2/\Lambda^2}}{1-\sqrt{1-2M^2/\Lambda^2}} \right)
    - \frac{\sqrt{1-2M^2/\Lambda^2}}{4\pi^2} ~ \right]^{-1} \label{zp}
\EQ
is related to normalization of a pseudoscalar state [see 
Eq.~(\ref{norm_fact})].
Clearly $m_\pi$ vanishes in the chiral limit $m_0\rightarrow 0$
 and the pseudoscalar state is identified with the Nambu-Goldstone boson.
In Eq.~(\ref{LFBSE_pion}), the first term corresponds to a kinetic energy part
  of the fermion and antifermion with the constituent mass $M$ and the 
  second term, a potential energy part.
The masslessness of the pseudoscalar state in the chiral limit is realized 
 by the exact cancellation between the kinetic energy and the potential energy.
Indeed, if we integrate Eq.~(\ref{LFBSE_pion}) over $x$ and $k_\perp^i$, 
 we find 
\BQA
m_\pi^2 \int_0^1 dx \int d^2k_\perp \phi_\pi (x,k_\perp^i)
&=&\left(1-\frac{2g^2_0\int_0^1 \frac{dx}{x} \int\frac{d^2q_\perp}{(2\pi)^3}}
           {\frac{m_0}{M}+2g^2_0\int_0^1\frac{dx}{x}\int\frac{d^2q_\perp}{(2\pi)^3}}
\right)
\int_0^1 dy  \int d^2 \ell_\perp \frac{\ell_\perp^2 + M^2}{y(1-y)}
    \phi_\pi (y , \ell_\perp^i)
\NN
&\rightarrow& 0\quad (m_0\rightarrow 0)~.
\EQA
Therefore,  $m_\pi=0$ is fulfilled in the chiral limit even though 
  the fermion has the constituent mass.

On the other hand, squared mass of the scalar state for small $m_0$ is
\BQ
 m_\sigma^2 = 4M^2 + {\cal O}(m_0)~.
\EQ
At a first glance, the result $m_\sigma=2M$ in the chiral limit 
  seems to suggest a static picture 
  of a fermion and an antifermion, but actually the mass of the scalar meson 
  comes from a part of the potential energy.
The kinetic energy cancels with the rest of the potential energy.

Equations~(\ref{solpb}) and (\ref{solsb}) have the same functional 
  form with respect to the variables $x$ and $k_\perp^i$.
However, the difference between $m_\pi$ and $m_\sigma$ greatly 
  affects the shape of the LC wavefunctions.
This is most remarkable in the chiral limit: 
As $m_0\rightarrow 0$, Eq.~(\ref{solpb}) becomes independent of $x$:
\BQA
&&\phi_\pi (x, k_\perp^i)
   \rightarrow -i \sqrt{N} \sqrt{Z_\pi}  \frac{1}{k_\perp^2 + M^2}~,
\EQA
where the constant $C_\pi$ was evaluated as 
$
  C_\pi \rightarrow  -i (2\pi)^3 (NZ_\pi)^{-1/2}.
$
On the other hand,  Eq.~(\ref{solsb})  shows a narrow peak at $x=1/2$.
Therefore, the pseudoscalar state is a highly collective state, 
  while the scalar state shows an approximate constituent picture.

Now let us compare our result Eq.~(\ref{solpb}) with those of the literatures 
 \cite{Heinzl_review,Bentz}.
First of all, equivalence with the result of Ref.~\cite{Bentz} is easily verified.
As we commented before, the unfamiliar spinor structure in 
  Eq.~(\ref{pionic_state}) is due to our specific choice of 
  the mode expansion and the representation for the $\gamma$ matrices.
If one uses the following mode expansion for the good 
  component of the fermion:
$$
\psi_+(x)=\sum_\lambda \int_{-\infty}^\infty \frac{d^2k_\perp}{2\pi}
\int_0^\infty \frac{dk^+}{\sqrt{2\pi k^+}}
\left[ \tilde b(\k,\lambda)u_+(\k,\lambda)e^{i\k\x}
+\tilde d^\dag(\k,\lambda)v_+(\k,\lambda)e^{-i\k\x} \right],
$$ 
one can obtain the same spinor structure as that of Ref.~\cite{Bentz}.
Of course the two LC wavefunctions should coincide with each other for
 observable quantities.
Indeed, both give the same (quark) distribution function 
  $q(x)=\int d^2k_\perp/(2\pi)^3 
  \sum_{\alpha,\beta}|\Phi_{\alpha\beta}(\k)|^2$ .

On the other hand, the result of Ref.~\cite{Heinzl_review} seems 
  different from ours Eq.~(\ref{solpb}).
The possible origin of the discrepancy might be attributed to the following 
  two points.
First of all, the author of Ref.~\cite{Heinzl_review} considered 
  the Melosh transformation~\cite{Melosh} which relates the LF spinor 
  and the usual spinor in the equal-time quantization.
Such nonstatic spin effects might be important when we discuss   
  phenomenological aspects of light mesons (for example, see 
  Ref.~\cite{Mankiewicz}). 
However, even if we take it into account, it is hard to see the coincidence.
Secondary but most importantly, he derived the pion LC wavefunction by 
  projecting the Bethe-Salpeter amplitude on the equal LC time plane. 
Though this procedure should give the same result as that of the LF 
  bound-state equation as far as we are considering only the ladder 
  ($1/N$ leading) contribution, equivalence of the two is a highly
  nontrivial problem in our complicated analysis.

\subsection{The Gell-Mann--Oakes--Renner and PCAC relations}

Now that we have the LC wavefunction for the pseudoscalar meson, it is 
 straightforward to obtain the decay constant $f_\pi$:
\begin{equation}
 i P^\mu f_\pi = \bra{0} j^\mu _5 (0) \ket{ ~\pi;  \itP}.
\end{equation}
For actual calculation, it is safer and easier to treat the plus component.
If we use the extended parity-invariant cutoff, the result is 
\BQ
f_\pi = 2M Z_\pi^{-\frac12} + {\cal O}(N^0)~.
\label{decay_const}
\EQ
Together with the pseudoscalar mass (\ref{pion_mass}) in the chiral limit, 
 we find the Gell-Mann, Oakes, and Renner relation,
\BQA
 m_\pi^2 f_\pi^2 &=& -4m_0 \left( - \frac{NM}{g_0^2} \right)    \nonumber \\
  &=& -4m_0 \bra{0} \bar \Psi \Psi \ket{0}~.
\EQA

The PCAC relation is also checked by using the state $\ket{~\pi ; \itP}$.
If we normalize the pseudoscalar field 
  $\pi_{\rm n}(x)\propto \bar \Psi(x) i \gfive \Psi(x) $ as
\BQ
 \bra{0} \pi_{\rm n}(0) \ket{~\pi ; \itP} = 1~,
\EQ
we find that $Z_\pi^{-1/2}$ given in Eq.~(\ref{zp}) is the normalization factor 
\BQ
 \pi_{\rm n}(x) = Z_\pi^{-\frac12}g^2
   \bar \Psi(x) i \gamma_5 \Psi(x) ~,\label{norm_fact}
\EQ
where we have used the gap equation.
Therefore, we arrive at the PCAC relation
\BQA
\partial_\mu j^\mu_5 &=& 2m_0  \bar \Psi(x) i \gamma_5 \Psi(x)  \nonumber \\
 &=& m_\pi^2  f_\pi \pi_{\rm n}(x)~.
\EQA
Note that the decay constant (\ref{decay_const}) and the normalization factor 
 (\ref{norm_fact}) are equivalent to the previous results (Eqs.~(5.25) and 
 (5.28) in Ref.~\cite{I}) in the infinitely heavy mass limit of bosons
  $\mu \rightarrow \infty $.

\subsection{Symmetric phase}

Here we consider the symmetric phase in the chiral limit $m_0=0$.
When $g^2_0 < g^2_{\rm cr}=2\pi^2/\Lambda^2$, the gap equation 
 (\ref{gap_NJL}) has only a trivial solution $M=0$.
A quantity which should be zero in the broken phase is 
 now estimated as 
\BQ
1-g^2_0\int \frac{d^3\k}{(2\pi)^3}\frac{\epsilon(k^+)}{k^+}
=1-\frac{g_0^2}{g_{\rm cr}^2}\neq 0~.
\EQ
Subsequently the factor $\alpha$ defined by Eq.~(\ref{alpha})
  is different from that of the broken phase [Eq.~(\ref{alpha_br})],
\BQ
\alpha^{-1}\equiv \alpha_{\rm sym}^{-1}= 
1-\frac{g_0^2}{g_{\rm cr}^2}+ \frac{2g_0^2}{(2\pi)^3}
\int_0^1 \frac{dx}{x} \int d^2 q_\perp        ~.
\EQ
Then, both of the LF bound-state equation for the scalar and pseudoscalar
 states are given by
\BQ
 m^2_{\rm sym} \phi_{\rm sym}(x , k_\perp^i)
 = \frac{k_\perp^2}{x(1-x)}\phi_{\rm sym}(x , k_\perp^i)
 - \frac{g_0^2 \alpha_{\rm sym}}{(2\pi)^3}\frac{1}{x(1-x)}
   \int_0^1 dy \int d^2 \ell_\perp \frac{\ell_\perp^2 }{y(1-y)}
   \phi_{\rm sym}(y, \ell_\perp^i)~.
\EQ
The solution to the bound-state equation is 
\BQ
\phi_{\rm sym}(x, k_\perp^i)=-C_{\rm sym}\frac{g_0^2}{(2\pi)^3} 
    \left(1-\frac{g_0^2}{g_{\rm cr}^2} \right)^{-1}
    \frac{1}{x(1-x)- k_\perp^2 / m_{\rm sym}^2}~,\label{sol}
\EQ
where $C_{\rm sym}$ is a normalization constant and 
  $m_{\rm sym}=m_\pi=m_\sigma$ is given as a solution of the equation
\BQ
 \frac{1}{g_0^2}- \frac{1}{g_{\rm cr}^2}
  =  \int_{0}^{1} dx \int \frac{d^2 k_\perp}{(2\pi)^3}
     \frac{ m_{\rm sym}^2}{ k_\perp^2  - m_{\rm sym}^2 x(1-x) }~.
\EQ
Again this is equal to the previous result of the chiral Yukawa model
   with $\mu^2 \rightarrow \infty$ (Eq.~(5.26) in Ref.~\cite{I})
   and therefore if we use the same cutoff as before, we obtain the same 
   result for $m_{\rm sym}$.
Moreover, though the above calculation was intended only to 
   $g_0^2<g_{\rm cr}^2$ case, if we increase the coupling constant over 
   its critical value $g_{\rm cr}^2$, we find a negative solution 
   $m_{\rm sym}^2<0$.
This implies that the symmetric solution causes instability when 
   $g_0^2>g_{\rm cr}^2$ and thus we must choose the broken solution.

\setcounter{equation}{0}
%
%
\section{Summary and Conclusion}

We have investigated a description of D$\chi$SB on the LF in the NJL model.
The importance of solving the fermionic constraint for the bad spinor 
  component was greatly stressed in analogy with the zero-mode constraint 
  of scalar models.
The exact classical solution enabled us to check the properties of the 
  LF chiral transformation.
Though the chiral transformation is differently introduced on the LF,
  we finally found the equivalence to the usual chiral transformation.

For a description of D$\chi$SB of LF NJL model, 
  it was very important to solve the fermionic constraint nonperturbatively 
  in quantum treatment.
To do so, we introduced a bilocal formulation and solved the bilocal fermionic
  constraint with a fixed operator ordering by the $1/N$ expansion.
Systematic $1/N$ expansion of the fermion bilocal operator is realized by
  the boson expansion method as the Holstein-Primakoff expansion. 
The leading bilocal fermionic constraint was identified with the gap 
  equation for the chiral condensate after we took care of the infrared 
  divergence.
If we choose a nontrivial solution of the gap equation, 
  we have a Hamiltonian in the broken phase but with a trivial vacuum.

The physical role of the fermionic constraint in the LF NJL model is very 
  similar to that of the zero-mode constraint for scalar models.
We have seen a close parallel between these two constraints.
Especially it should be noted that the gap equation came from the longitudinal 
  zero mode of the bilocal fermionic constraint.

It is very natural that we can reach the broken phase by solving 
  the quantum fermionic constraint by $1/N$ expansion
  because the fermionic constraint is originally a part of the 
  Euler-Lagrange equation and thus must include relevant information 
  of dynamics.
What we did is very similar to the usual mean-field approximation for the 
  Euler-Lagrange equations.
Indeed the leading order in the $1/N$ expansion corresponds to the 
  mean-field approximation.
However, our way of solving the fermionic constraint with the boson 
  expansion method can easily go beyond the mean-field level.
Such higher order calculation enabled us to derive a correct broken
  Hamiltonian and to show the divergent behavior of the (spatial 
  integration of) pseudoscalar field.

Independence of mass from the mode expansion has both desirable and 
  undesirable aspects.
The inclusion of correct mass dependence into the IR divergent integral
  was required when we identify the lowest fermionic constraint with 
  the gap equation.
This is the point we must always take care of.
On the other hand, the Fock vacuum is defined independent of the value of 
  mass. 
Due to this fact, it is enough to have only one vacuum, namely, the Fock
  vacuum even in the chirally broken phase.
This is the favorable aspect.
However, the cost of such a simple vacuum was payed by, for example, 
   unusual chiral transformation of fields such as $[Q_5^{\rm LF}, 
    \bar\Psi i\gfive\Psi]\neq -2i \bar\Psi\Psi$ 
   and non-vanishing of the LF chiral charge $[Q_5^{\rm LF}, H]\neq 0$ 
   in the broken phase.
We found that the both effects are proportional to the dynamical fermion 
   mass $M$.
We also insisted the necessity of a bare mass term which accurately produced
   the constituent mass term.
Although the special role of the fermionic constraint might be restricted
   to the LF NJL model, the unusual chiral transformation and the 
   nonconservation of the chiral charge are general features of the 
   chiral symmetry breaking on the LF.
This is because they are natural consequences of the
  coexistence of the chiral symmetry breaking and the Fock vacuum.

The leading order eigenvalue equation for a single bosonic state is 
  equivalent to the leading order fermion-antifermion bound-state equation. 
The bound-state equations were solved analytically for scalar and 
  pseudoscalar mesons and we obtained their light-cone wavefunctions 
  and masses.
The meson masses, the decay constant, and so on were fairly consistent 
  with those of our previous analysis on the chiral Yukawa model.
The leading order calculation was limited only to two body sector 
  (fermion and antifermion).
If we consider the higher order Hamiltonian such as $h^{(3)}$ or $h^{(4)}$, 
  we will be able to discuss four or six body sectors.
In other words, since we have bosonic meson states, we can expand the Fock 
  space in terms of the mesonic degrees of freedom.
Then, for example, we will be able to discuss the mixing of scalar state and 
  two pseudoscalar fields ($\pi$-$\pi$ mixing with $\sigma$).

\section*{Acknowledgments}

The authors acknowledge W. Bentz for discussions on the cutoff schemes.
One of them (K.I.) is thankful to K. Harada and T. Heinzl for helpful
  discussions at the very early stage of this work and to K. Yazaki 
  and K. Yamawaki for their useful comments.
He is also grateful to T. Hatsuda and H. Toki for their encouragements
  and interests in our work.

%
%
\appendix
\setcounter{equation}{0}
\section{Conventions}

We follow  the Kogut-Soper convention~\cite{Review}.
First of all, the light-front coordinates are defined as
\begin{equation}
x^\pm = \frac{1}{\sqrt2}(x^0\pm x^3), \quad
x^i_\perp = x^i \quad(i=1,2), 
\end{equation}
where we treat $x^+$ as ``time''. 
The spatial coordinates $x^-$ and $x_\perp$ 
  are called the longitudinal and transverse 
  directions respectively. 
Derivatives in terms of $x^\pm$ are defined by
$\partial_\pm=\partial/\partial x^\pm$.
It is useful to introduce projection operators $\Lambda_\pm$ 
  defined by
\begin{equation}
\Lambda_{\pm}=\frac12 \gamma^\mp \gamma^\pm
=\frac{1}{\sqrt2}\gamma^0\gamma^\pm.
\end{equation}
Indeed $\Lambda_\pm$ satisfy the projection properties 
$\Lambda_\pm^2=\Lambda_\pm,\ \Lambda_++\Lambda_-=1$, etc.
Splitting the fermion field by the projectors as
\begin{equation}
  \Psi^a = \psi^a_+ + \psi^a_- , \quad 
  \psi^a_{\pm} \equiv \Lambda_{\pm}\Psi^a ,
\end{equation}
we find that for any fermion on the LF, 
  $\psi_-$ component is a dependent degree of freedom. 
$\psi_+$ and $\psi_-$ are called the ``good component'' and 
  the ``bad component'', respectively.

As is commented in the text, for practical calculation, we use 
  the two-component representation for the gamma matrices.
The two-component representation is characterized by a specific 
  form of the projectors $(\ref{two-compo})$. 
Then the projected fermions $\psi_\pm$ have only two components.
There are many possibilities which realize Eq.~$(\ref{two-compo})$.
For example, a specific representation 
\BQ
\gamma^0=\pmatrix{
0 & -i\cr
i&0
},\quad
\gamma^3=
\pmatrix{
0& i\cr
i&0
}, \quad
\gamma^i=
\pmatrix{
-i\sigma^i&0\cr
0&i\sigma^i
}
\EQ
is used in Ref.~\cite{Hari-Zhang}.
In this paper, however, we choose a representation (\ref{two-compo-gamma}) 
  from which it is easy to extract information of the 1+1 dimensional results.
Two-component spinors $\psi$ and $\chi$ are defined by Eq.~(\ref{spinors}). 
Results of the chiral Gross-Neveu model can be 
  easily obtained if we make a replacement for the Pauli matrices 
  $\sigma_3 \rightarrow 1$ and $\sigma^i\rightarrow 0$, 
  and regard $\psi$ and $\chi$ as one component spinors. 

Using this representation, the Lagrangian density of the NJL model 
  is written as
\BQA
{\cal L}&=&i \psi^\dagger \del_+\psi + i\chi^\dagger \del_-\chi
-\frac{1}{\sqrt{2}}\left( \psi^\dagger \sigma^i\del_i\chi
-\chi^\dagger \sigma^i\del_i\psi   \right) 
-\frac{m_0}{\sqrt{2}}\left(\psi^\dagger\chi + \chi^\dagger\psi \right)\NN
&&+\frac{g^2}{4}\left\{
(\psi^\dagger\chi + \chi^\dagger\psi)^2
-(\psi^\dagger\sigma^3\chi - \chi^\dagger\sigma^3\psi)^2
\right\}.
\EQA

\setcounter{equation}{0}
\section{Classical Solutions to the fermionic constraints}
To solve the fermionic constraints $classically$ means that 
  we treat all the fermion fields (both good and bad components) 
  as Grassmann numbers and neglect all the $c$-numbers which will 
  emerge in quantum theory under the exchange of variables.

Before discussing a complicated equation of the NJL model, it would be 
  better to go first with the chiral Gross-Neveu model.
We solve the fermionic constraint of the chiral Gross-Neveu model 
with the antiperiodic boundary condition:
\BQA
&& \Big\{i\del_- + g^2 a(x^-) \Big\}\chi =\frac{m_0}{\sqrt2}~\psi ~,\\
&&\chi_a(x^-=-\infty)=-\chi_a(x^-=\infty)~,
\EQA
where we used a matrix notation with a matrix 
  $a_{ab}(x^-) \equiv \psi_a(x^-)\psi_b^\dagger(x^-)$.
The solution to this equation is given by 
\BQ
\chi(x^-)=\int_{-\infty}^\infty dy^- G_{\rm GN}(x^-, y^-)\frac{m_0}{\sqrt2}\ \psi(y^-)~,
\label{solution_CGN}
\EQ
where $G_{\rm GN}(x^-,y^-)$ is the Green function satisfying
\BQA
&&\left\{ i\del_-^x + g^2 a(x^-)\right\} G_{\rm GN}(x^-,y^-)
  =\delta(x^--y^-)~,\label{Green_eq}\\ 
&&G_{\rm GN}(x^-=-\infty, y^-)=-G_{\rm GN}(x^-=\infty, y^-)~.\label{bc_G}
\EQA
Due to Eq.~(\ref{bc_G}), the solution of course satisfies 
  the antiperiodic boundary condition.
Equation~(\ref{Green_eq}) is solved as 
\BQA
G_{\rm GN}(x^-, y^-)&=&G_{\rm GN}^{(0)}(x^-)
              \left[~\frac{1}{2i}\epsilon(x^--y^-) + C ~\right]
                G_{\rm GN}^{(0)-1}(y^-)~,
           \label{Green_CGN}\\
&&G_{\rm GN}^{(0)}(x^-)={\rm P}\ e^{ig^2\int_{-\infty}^{x^-}a(y^-)dy^-}~,\\
&&C=-\frac{1}{2i}\frac{G_{\rm GN}^{(0)}(\infty)-G_{\rm GN}^{(0)}(-\infty)}
    {G_{\rm GN}^{(0)}(\infty)+G_{\rm GN}^{(0)}(-\infty)} ~,
\EQA
where $G_{\rm GN}^{(0)}(x^-)$ is a solution of a homogeneous equation 
  $\left\{i\del_-^x+ g^2 a(x^-) \right\} G_{\rm GN}^{(0)}(x^-)=0$ 
  and the integral constant $C$ has been determined so that 
  $G_{\rm GN}(x^-,y^-)$ satisfies the antiperiodic boundary condition.
When $N=1$, the solution (\ref{solution_CGN}) is equivalent to 
  Domokos' solution to the Thirring model on the light front~\cite{Domokos}.

In two dimensions, the LF chiral transformation is not distinguishable 
  with the $U(1)$ transformation. 
Indeed, the ``LF chiral transformation'' on the good component is 
  $\psi \rightarrow e^{i\theta}\psi$ and equivalent to the $U(1)$ 
  transformation. 
And also, the solution (\ref{solution_CGN}) implies that the 
  bad component rotates just the same way as the good component 
  $\chi \rightarrow e^{i\theta}\chi$.

Next let us turn to the NJL model.
If we explicitly write all the indices, the fermionic constraint (\ref{FC}) is 
\BQ
i\del_-\pmatrix{\chi_{1a}\cr \chi_{2a}}
=\frac{1}{\sqrt2}\pmatrix{
m_0\psi_{1a}-\del_z\psi_{2a}\cr
-\del_{\bar z}\psi_{1a}+m_0\psi_{2a}
}-g^2\pmatrix{
\psi_{1a}\psi_{1b}^\dagger\ \chi_{1b}-\psi_{1a}\psi_{2b}\ \chi_{2b}^\dagger\cr
\psi_{2a}\psi_{2b}^\dagger\ \chi_{2b}-\psi_{2a}\psi_{1b}\ \chi_{1b}^\dagger\cr
}~,
\EQ
where $\del_z=\del_1-i\del_2$ and $\del_{\bar z}=\del_1+i\del_2$.
Since the equation for $\chi_1$ (or $\chi_2$) includes $\chi_1$ and 
  $\chi_2^\dag$ (or $\chi_2$ and $\chi_1^\dag$), it is useful to 
  introduce a constraint equation for $-\chi_2^\dagger$  instead of $\chi_2$. 
Then we have a more tractable equation
\BQ
i\del_-\pmatrix{
\chi_{1a}\cr
-\chi_{2a}^\dagger
}=\frac{1}{\sqrt2}\pmatrix{
m_0\psi_{1a}-\del_z\psi_{2a}\cr
-\del_z\psi_{1a}^\dagger+m_0\psi_{2a}^\dagger
}
-g^2\pmatrix{
\psi_{1a}\psi_{1b}^\dagger & \psi_{1a}\psi_{2b} \cr
\psi_{2a}^\dagger\psi_{1b}^\dagger & \psi_{2a}^\dagger \psi_{2b} \cr
}
\pmatrix{
\chi_{1b}\cr
-\chi_{2b}^\dagger
}~.
\EQ
As in the 1+1 dimensional case, the solution is immediately given 
  if we find the Green function $G(x^-,y^-,x_\perp)$ which satisfies
\BQ
\Big\{i\del_-^x+g^2 {\cal A}(x^-)\Big\}G(x^-,y^-,x_\perp)=\delta(x^--y^-)~,
\EQ
with a matrix ${\cal A}_{ijab}(x)$ defined by Eq.~(\ref{matrix_A}).
The result is very similar to the two-dimensional result and is 
  given by Eqs.~(\ref{classical_solution}) and (\ref{Green_NJL}) in the text.

\setcounter{equation}{0}
\section{Problem of the operator ordering}

Here we consider  the  problem of operator ordering within 
  the chiral Gross-Neveu model with $N=1$.
Following the standard procedure, the Dirac brackets are calculated as
\BQA
&&\left\{\psi(x), \psi^\dagger(y)\right\}_{\rm D}=-i\delta(x^- -y^-)~,\\
&&\left\{\chi(x), \psi^\dagger(y)\right\}_{\rm D}= -i G_{\rm GN}(x,y) 
\left(\frac{m_0}{\sqrt2}-g^2\psi^\dagger(y)\chi(y)\right)~,\label{DB}\\
&&\left\{\chi(x), \psi(y)\right\}_{\rm D}=-iG_{\rm GN}(x,y) g^2 \psi\chi(y)~,
\EQA
where $G_{\rm GN}(x,y)$ is the Green function (\ref{Green_CGN}) for $N=1$ 
  case.
To quantize the system we simply replace the Dirac bracket  
  $\{A,B\}_{\rm D}$ by the anticommutation relation $-i \{A,B\}$. 
This procedure has ambiguity of the operator ordering.

The operator ordering we took for the fermionic constraint (\ref{FC})
  in the NJL model corresponds to the following one in the chiral 
  Gross-Neveu model,
\BQ
i\del_-\chi + g^2\psi\psi^\dagger\chi =\frac{m_0}{\sqrt2}\psi~.\label{fc_cgn}
\EQ
We can easily find its {\it quantum} solution due to 
  $[\psi\psi^\dag(x),\psi\psi^\dag(y)]=0$.
The solution is 
\BQ
\chi_{\rm sol}(x^-)
=\int_{-\infty}^\infty dy^-\ G_{\rm GN}(x,y)\frac{m_0}{\sqrt2}\psi(y)~,\label{sol_cgn}
\EQ 
where $G$ is again the Green function (\ref{Green_CGN}) with $N=1$.

Now let us consider the consistency for the anticommutator 
  $\{\chi, \psi^\dag\}$.
It can be calculated two different ways:
(i) from the solution $\chi_{\rm sol}$ of the fermionic constraint, 
and (ii) from the Dirac bracket (\ref{DB}).
We fix the operator ordering of the fermionic constraint by Eq.~(\ref{fc_cgn})
 and check whether the Dirac bracket can produce the same 
 anticommutator or not.

Instead of the anticommutator itself, we present here the calculation of 
  a quantity $iD_-^x \{\chi(x), \psi^\dagger(y)\}$ where 
 $iD_-^x=i\del_-+g^2\psi\psi^\dag$.
Using the solution (\ref{sol_cgn}), we have 
\BQ
iD_-^x\left\{\chi_{\rm sol}(x), \psi^\dagger(y)\right\}
=\delta(x^--y^-)\left(\frac{m_0}{\sqrt2}-g^2\psi^\dagger\chi
\right)~.\label{result-1}
\EQ
On the other hand, if we take the simplest ordering in 
the r.h.s. of the Dirac bracket (\ref{DB}), we obtain 
\BQA
iD_-^x\left\{\chi(x), \psi^\dagger(y)\right\}&=&
iD_-^xG_{\rm GN}(x,y)\left(\frac{m_0}{\sqrt2}-g^2\psi^\dagger\chi(y)\right)\NN
&=&\delta(x^--y^-)\left(\frac{m_0}{\sqrt2}-g^2\psi^\dagger\chi(y)\right)~.
\EQA
This is identical with the result (\ref{result-1}).
Therefore we find our ordering Eq.~(\ref{fc_cgn}) is consistent 
  with the anticommutation relation 
\BQ
\left\{\chi(x), \psi^\dagger(y)\right\}
=G_{\rm GN}(x,y)\left(\frac{m_0}{\sqrt2}-g^2\psi^\dagger\chi
\right)~.
\EQ
Of course if we take other operator ordering, the two results do not coincide.
We expect that even in the NJL model, we can select the r.h.s. of 
  Dirac brackets so that they coincide with the direct result with the 
  ordering defined by Eq.~(\ref{FC}).

\setcounter{equation}{0}
\section{Bilocal Fermionic constraints and their solutions by the 
         Boson Expansion Method}

It is tractable to solve the equations for $\T(\x,\y)$ and $\U(\x,\y)$
  rather than $\T_{\alpha\beta}(\x,\y)$ and $\U_{\alpha\beta}(\x,\y)$.
In momentum representation, the fermionic constraints for $\T$ and $\U$ are 
\BQA
q^+ \T(\p,\q)&=&\frac12 \left(-iq_\perp^i \sigma^i +m_0\right)_{\alpha\beta}
               \Big(\M_{\alpha\beta}(\p,\q)-\M_{\alpha\beta}(\q, \p)\Big)\NN
            &&-\frac{g^2}{2}\int \frac{d^3\p'd^3\q'}{(2\pi)^3}
              \left\{\M_{\alpha\beta}(\p,\q-\p'-\q')
                     \left( \delta_{\alpha\beta}\T(\p',\q')
                     + \sigma^3_{\alpha\beta}i~\U(\p',\q') \right) \right. \NN
            &&\qquad \left. - \left( \delta_{\alpha\beta}\T(\p',\q')
                       - \sigma^3_{\alpha\beta}i~\U(\p',\q')  \right) 
                   \M_{\alpha\beta}(\q-\p'-\q', \p)   \right\}~,\label{FCT}\\  
q^+ i~\U(\p,\q)&=&\frac12 \left[
              \left\{\sigma^3(-iq_\perp^i \sigma^i +m_0)\right\}_{\alpha\beta}
                 \M_{\alpha\beta}(\p,\q)
             +\left\{(-iq_\perp^i \sigma^i +m_0)\sigma^3\right\}_{\alpha\beta}
                 \M_{\alpha\beta}(\q, \p)\right]\NN
            &&-\frac{g^2}{2}\int \frac{d^3\p'd^3\q'}{(2\pi)^3}
              \left\{ \M_{\alpha\beta}(\p,\q-\p'-\q')
                     \left( \sigma^3_{\alpha\beta}\T(\p',\q')
                    + \delta_{\alpha\beta}i~\U(\p',\q')  \right) \right. \NN
            &&\qquad \left.+ \left( \sigma^3_{\alpha\beta}\T(\p',\q')
                       - \delta_{\alpha\beta}i~\U(\p',\q')  \right) 
                     \M_{\alpha\beta}(\q-\p'-\q', \p)   \right\} ~.\label{FCU}
\EQA
In place of the quantization condition (\ref{Quantization}), 
   the system with bilocal operators can be characterized by 
   the following algebra
\BQA
&&\hspace{-1.5cm}
\Big[:\M_{\alpha\beta}({\p}_1,{\p}_2):\ ,\ \  
     :\M_{\gamma\delta}({\q}_1,{\q}_2):\Big]\NN
&=& :\M_{\alpha\delta}({\p}_1,{\q}_2):
      \delta_{\beta\gamma}\delta^{(3)}({\p}_2+{\q}_1)\ 
  - :\M_{\gamma\beta}({\q}_1,{\p}_2):
  \delta_{\alpha\delta}\delta^{(3)}({\p}_1+{\q}_2) \nonumber \\
&&+N\delta_{\alpha\delta}\delta_{\beta\gamma}
    \delta^{(3)}({\p}_1+{\q}_2)\delta^{(3)}({\p}_2+{\q}_1)\NN 
&&\quad \times \Big(\theta(p_1^+)\theta(p^+_{2})
                    \theta(-q^+_{1})\theta(-q^+_{2})
                  - \theta(-p^+_{1})\theta(-p^+_{2})
                    \theta(q^+_{1})\theta(q^+_{2})\Big)~,
\label{algebra}
\EQA
where the normal order of $\M$ was defined with respect to the Fock 
  vacuum (\ref{vac}) 
\BQ
   :\M_{\alpha\beta}^{+-}(\p,\q):\ket{0}
 =\ :\M_{\alpha\beta}^{-+}(\p,\q):\ket{0} 
 =\ :\M_{\alpha\beta}^{++}(\p,\q):\ket{0}=0~.
\EQ
The upper indices stand for signs of the longitudinal momenta.

The complicated structure of the algebra for the bilocal operators 
  which originates from the fermion statistics, 
  is greatly reduced if one introduces the boson expansion method. 
We can represent the operators $:\M:$ in terms of bilocal boson operators
  $B(\p,\q)$ of order ${\cal O}(N^0)$ so that they fulfill the original 
  algebra (\ref{algebra}).
Since the algebra has a bosonic feature in the large $N$ limit,  
$$
\Big[:\M_{\alpha\beta}^{++}({\p}_1,{\p}_2):\ ,\   
     :\M_{\gamma\delta}^{--}({\q}_1,{\q}_2):\Big]
\longrightarrow N \delta_{\alpha\delta}\delta_{\beta\gamma}
    \delta^{(3)}({\p}_1+{\q}_2)\delta^{(3)}({\p}_2+{\q}_1)~,
$$
it would be better to choose a representation which satisfies 
  this\footnote{Actually there are many possibilities to express 
  Eq.~(\ref{algebra}) in terms of bosonic operators, corresponding to various 
 ``local expansions'' of the Grassmannian manifold of the bilocal operators.}.
The Holstein-Primakoff type expansion satisfies the requirement.

Physically this procedure corresponds to extracting purely bosonic degrees 
  of freedom from a fermion-antifermion system, i.e., a mesonic system.
The power of the boson expansion method in the light-front field theories 
  was first demonstrated by one of the authors~\cite{BEM_itakura}.
He applied the Holstein-Primakoff type expansion to 1+1 dimensional 
  QCD and derived an effective Hamiltonian for mesons as local bosons
  whose masses are given by the 't~Hooft equation.
Using the effective Hamiltonian, we can in principle study, say, 
  scattering of mesons as $q\bar q$ bound states.

Since the essential structure of the algebra (\ref{algebra}) is 
   determined only by the longitudinal momentum, 
   it is straightforward to apply the Holstein-Primakoff
   expansion discussed in Ref.~\cite{BEM_itakura} to four dimensional 
   fermionic theories.
Indeed the operators $:\M:$ are represented as follows:
\BQA
:\M^{-+}_{\alpha\beta}(\p_1,\p_2):
  &=& \int [d\q] \sum_{\gamma} 
      B^\dagger_{\alpha\gamma} (-\p_1,\q) B_{\beta\gamma}(\p_2,\q) 
      \equiv A_{\beta\alpha}(\p_2,-\p_1)~, \label{BEM_-+}\\
:\M^{+-}_{\alpha\beta}(\p_1,\p_2):
  &=& -\int [d\q]\sum_{\gamma} 
      B^\dagger_{\gamma\beta}(\q,-\p_2) B_{\gamma\alpha}(\q,\p_1)~,
  \label{BEM_+-}\\
:\M^{++}_{\alpha\beta}(\p_1,\p_2):
  &=& \int [d\q]\sum_{\gamma}
     (\sqrt{N-A})_{\beta\gamma}(\p_2,\q)B_{\gamma\alpha}(\q,\p_1)~,
  \label{BEM_++}\\
:\M^{--}_{\alpha\beta}(\p_1,\p_2): 
  &=& \int [d\q]\sum_{\gamma}
      B^\dagger_{\gamma\beta} (\q,-\p_2)
      (\sqrt{N-A})_{\gamma\alpha}(\q,-\p_1)~.\label{BEM_--}
\EQA
These give the $1/N$ expansion of $\M_{\alpha\beta}(\p,\q)$.
The first few terms are shown in the text [Eqs.~(\ref{BEM_0})-(\ref{BEM_2})].

If we expand the bilocal operators $\T(\p,\q),\ \U(\p,\q)$, and  
  $\M_{\alpha\beta}(\p,\q)$,
  the equation for order $n$ can be written in a compact form:
\begin{equation}
\pmatrix{
t^{(n)}(\p,\q)\cr
u^{(n)}(\p,\q)}
= 
\pmatrix{
F^{(n)}(\p,\q)\cr
G^{(n)}(\p,\q)
}
-g_0^2 \frac{\epsilon(p^+)}{q^+}\int_{-\infty}^\infty
\frac{d^3\k}{(2\pi)^3}
\pmatrix{
   t^{(n)}(\k,\p+\q-\k)\cr
   u^{(n)}(\k,\p+\q-\k)\cr
}~,
\end{equation}
where quantities $F^{(n)}(\p,\q)$ and $G^{(n)}(\p,\q)$ are generally 
   complicated functions of bilocal operators 
   except for the lowest order [see Eq.~(\ref{LOFC})].
For example, $F^{(1)}$ and $G^{(1)}$ are 
\BQA
&& F^{(1)}(\p,\q) =\frac{1}{2q^+}(-i q^i_\perp\sigma^i + M )_{\alpha\beta}
   \left( \mu^{(1)}_{\alpha\beta}(\p,\q) - \mu^{(1)}_{\alpha\beta}(\q,\p)
\right)~,\\
&& G^{(1)}(\p,\q) = \frac{-i}{2q^+}\left[ 
\left\{\sigma_3(-i q^i_\perp \sigma^i + M )\right\}_{\alpha\beta}
   \mu^{(1)}_{\alpha\beta}(\p,\q) +
\left\{(-i q^i_\perp \sigma^i + M )\sigma_3\right\}_{\alpha\beta}
 \mu^{(1)}_{\alpha\beta}(\q,\p)
\right]~,
\EQA
where $\mu^{(1)}_{\alpha\beta}(\p,\q)$ is given by the boson expansion method 
 Eq.~(\ref{BEM_1}).
Since all of the orders of the operators are less than $n$, 
  we can solve this equation order by order.
The solution of this integral equation is 
\BQ
\pmatrix{
 t^{(n)}(\p,\q)\cr
 u^{(n)}(\p,\q)
}
= 
\pmatrix{
F^{(n)}(\p,\q)\cr 
G^{(n)}(\p,\q)
}- g_0^2 \frac{\epsilon(p^+)}{q^+}\alpha(p^++q^+)
    \int_{-\infty}^\infty \frac{d^3\k'}{(2\pi)^3} 
\pmatrix{
F^{(n)}(\k',\p+\q-\k') \cr
G^{(n)}(\k',\p+\q-\k') }~,
\EQ
where 
\BQ
\alpha(P^+)=\left(1+ g_0^2 \int_{-\infty}^\infty \frac{d^3 \k }{(2\pi)^3}
    \frac{\epsilon(k^+)}{P^+-k^+}\right)^{-1}~.\label{alpha}
\EQ
The quantities $t^{(2)}(\p,\q)$ and $u^{(2)}(\p,\q)$ are necessary 
  for obtaining a correct Hamiltonian of the system.

\end{document}